\documentclass[
    superscriptaddress,
    reprint,
    showkeys,
    showpacs,
    amsmath,
    amssymb,
    aps,
    prb,
    floatfix
]{revtex4-1}
\usepackage{physics}
\usepackage{siunitx}
\usepackage{graphicx}
\usepackage{dcolumn}
\usepackage{bm}
\usepackage{braket}
\usepackage[caption=false]{subfig}
\usepackage{float}
\usepackage{booktabs}
\usepackage{multirow}
\usepackage{tabularx}
\usepackage{color}
\definecolor{blue}{rgb}{0.3,0.3,0.9}
\usepackage{hyperref}
\hypersetup{
    colorlinks=true,
    linkcolor=blue,
    citecolor=blue,
    urlcolor=blue
}
\begin{document}
\title{Electric field exfoliation and high-T$_{\text{C}}$ superconductivity in field-effect hole-doped hydrogenated diamond (111)}
\author{D. Romanin}
\email{davide.romanin@polito.it}
\affiliation{
    Department of Applied Science and Technology, Politecnico di Torino, 10129 Torino, Italy
}
\author{Th. Sohier}
\affiliation{
    Theory and Simulation of Materials (THEOS), and National Centre for Computational Design and Discovery of Novel Materials (MARVEL), \'Ecole Polytechnique F\'ed\'erale de Lausanne, CH-1015 Lausanne, Switzerland
}
\author{D. Daghero}
\affiliation{
    Department of Applied Science and Technology, Politecnico di Torino, 10129 Torino, Italy
}
\author{F. Mauri}
\affiliation{
    Dipartimento di Fisica, Universit\'a di Roma La Sapienza, Piazzale Aldo Moro 5, I-00185 Roma, Italy and Graphene Labs, Fondazione Istituto Italiano di Tecnologia, Via Morego, I-16163 Genova, Italy
}
\author{R. S. Gonnelli}
\affiliation{
    Department of Applied Science and Technology, Politecnico di Torino, 10129 Torino, Italy
}
\author{M. Calandra}
\affiliation{
    Sorbonne Universit\'e, CNRS, Institut des Nanosciences de Paris, UMR7588, F-75252, Paris, France
    }

\date{\today}
\begin{abstract}
  We investigate the possible occurrence of field-effect induced superconductivity in the hydrogenated $(111)$ diamond surface by first-principles calculations. By computing the band alignment between bulk diamond and the  hydrogenated surface we show that the electric field exfoliates the sample, separating the electronic states at the valence band top from the bulk projected ones. At the hole doping values considered here, ranging from $n=2.84\times10^{13}$cm$^{-2}$  to $n=6\times 10^{14} $ cm$^{-2}$,
the valence band top is composed of up to three electronic bands hosting holes with different effective masses. These bands resemble those of the undoped surface, but they
are heavily modified by the electric field and differ substantially from a rigid doping picture. We calculate superconducting properties by including the effects of charging of the slab and of the electric
field on the structural properties, electronic structure, phonon dispersion and electron-phonon coupling. We find that at doping as
large as $n=6\times 10^{14} $ cm$^{-2}$, the electron-phonon interaction is $\lambda=0.81$ and superconductivity emerges with $T_{\text{C}}\approx 29-36$K.
Superconductivity is mostly supported by in-plane diamond phonon vibrations and to a lesser extent by some out-of-plane vibrations. The relevant electron-phonon scattering processes
involve both intra and interband scattering so that superconductivity is multiband in nature.
\end{abstract}


\maketitle

\section{\label{sec:intro}Introduction}

The metallization of 3D covalent solids can be beneficial for superconductivity, as the cases of MgB$_2$~\cite{MgB2}, H$_3$S~\cite{H3S} and B-doped diamond~\cite{Cohen} clearly show.
Indeed, covalent materials like diamond, boron nitride, silicon carbide are ultrahard and have highly energetic phonon frequencies, which is in principle beneficial for superconductivity, but are mostly insulating due to the nature of their covalent bonding (electrons are not mobile). Driving them to the superconducting state, however, requires the introduction of a sizeable number of carriers because of the slow increase of the density of states (DOS) as a function of doping.
An example is B-doped diamond that undergoes an insulator-to-metal transition at a boron concentration $n_B\approx10^{20}$~cm$^{-3}$ and then becomes a superconductor at $n_B\approx4-5\cdot10^{21}$~cm$^{-3}$ with a transition temperature $T_\text{C}=4$~K\cite{Ekimov, Bustarret}. The introduction of larger amount of dopants has proven to be problematic~\cite{Q-Carbon} because of the low solubility limit. It is possible to overcome the problem limit via non-equilibrium techniques as in \textrm{Q-carbon}~\cite{Q-Carbon2} or \textrm{Si}~\cite{Si}.

\begin{figure}
\centering
      \includegraphics[width=0.8\linewidth]{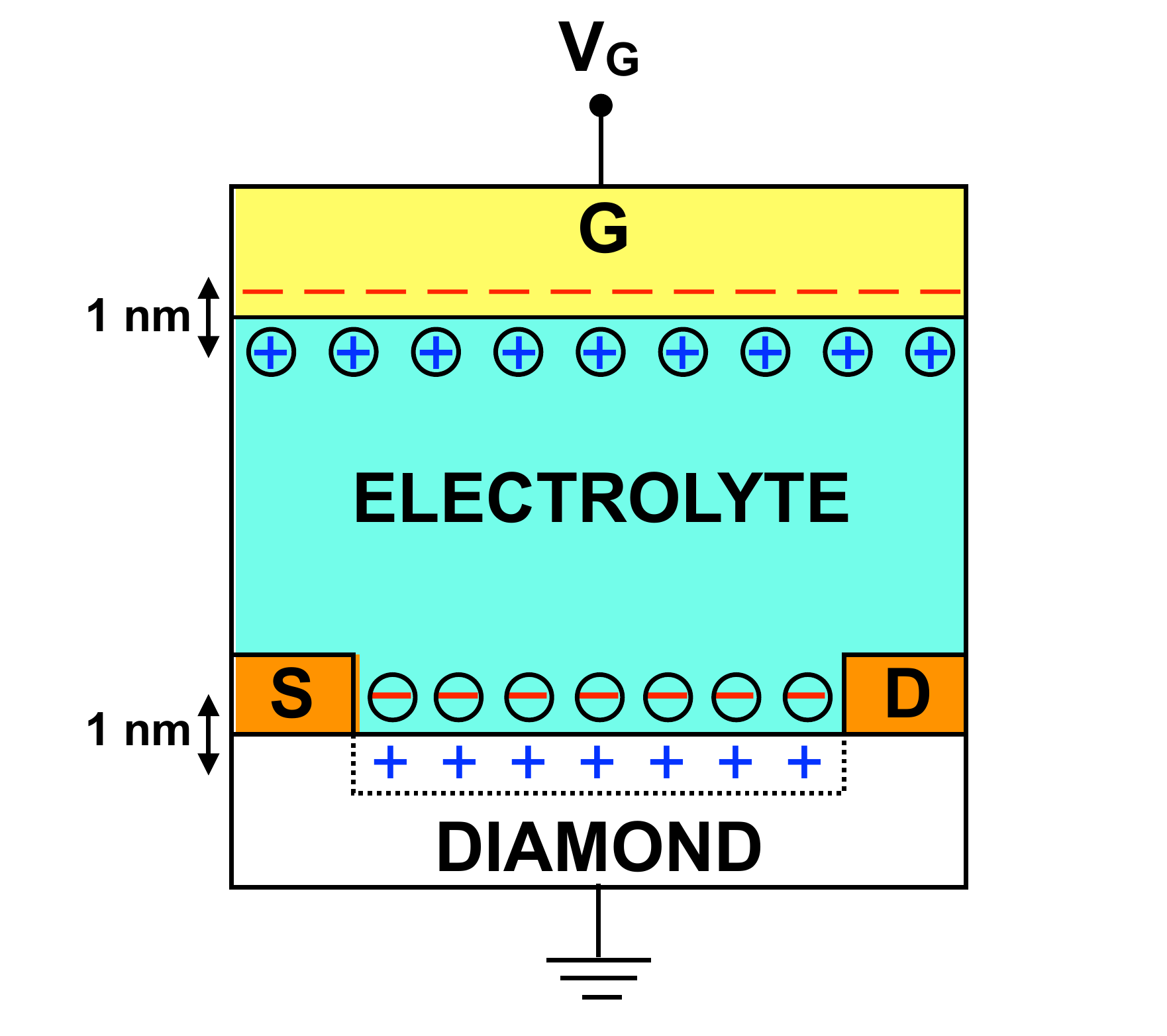}
      \caption
      {Field effect transistor (FET) geometry for the electrochemical gating. G, S and D are respectively the \emph{gate}, \emph{source} and \emph{drain} electrodes. $V_g$ is the gate tension.}
   \label{fig:el_gat}
\end{figure}

An alternative route to bypass the slow increase of the DOS in bulk samples is to dope 2D materials or surfaces where the density of states is constant as a function of energy. These systems are also appealing because doping techniques alternative to the chemical ones can be applied. For example, one way to avoid the introduction of boron in diamond samples is through the field effect, i.e. the induction of additional charges at the surface by means of a transverse electric field in a FET-like configuration. In a conventional FET, however, the applied electric field is limited by the breakdown field of the solid dielectric, and by the fact that the thickness of the latter must be large enough to ensure the absence of pinholes.
Electrochemical gating allows overcoming both these limitations. The technique consists in replacing the solid dielectric with an electrolyte (usually an ionic liquid or a polymer-electrolyte solution)  as shown in Fig.~\ref{fig:el_gat} for the case of diamond. When an electric voltage (for example, negative)  is applied to the gate electrode, ions are accumulated at the diamond surface forming a capacitor with a very small inter-plate distance. This results in a much larger density of induced charge carriers \cite{Gonnelli} (for example, holes) on the first few carbon layers.

Previous computations on the hydrogenated (110) diamond surface in FET configuration~\cite{Freeman,Sano} suggested a possible superconductive phase transition. However, in that case, the presence of the electric field was taken into account in a self-consistent way only for the electronic structure and not for vibrational properties. Actually the natural crystal facets of polycrystalline diamond films grown by chemical vapor deposition (CVD) are the (100) and (111) surfaces\cite{Ristein}. Moreover, experiments show that the surface capacitance of the hydrogenated (111) diamond is $2.6-4.6$ $\mu$F$/$cm$^2$ (Refs.~\onlinecite{Yamaguchi,Takahide}), while that of hydrogenated (100) diamond is $2.1-2.8$ $\mu$F$/$cm$^2$ (Refs.~\onlinecite{Yamaguchi,Akhgar,Takahide2}).
Therefore the (111) orientation of diamond allows accumulating larger charge densities at the surface and, consequently, obtaining a higher number of carriers at the Fermi level.

In this work we study the possibility of inducing superconductivity in (111) hydrogenated diamond thin films by means of electrochemical gating, using first-principles techniques. We consider the effect of hole doping and of the FET geometry in a self-consistent way both on electrons, phonons and on the electron-phonon interaction, by using a recently developed theoretical approach\cite{Thibault}.

The paper is organized as follow. In Sec.~\ref{sec:methods} we expose the computational details used for ab-initio calculations. After that, in  Sec.~\ref{sec:results} we move to the discussion of the results. In particular, in Sec.~\ref{sec:relax} we analyze the effects of the electric field on the relaxation of structural parameters, while  in Sec.~\ref{sec:exfoliation} we show that there is a field-driven `exfoliation' of the electronic bands. In Sec.~\ref{sec:charge} we investigate how the charge distribution is affected by the electrochemical gating and in Sec.~\ref{sec:elec} we study the electronic structure of our system by varying the amount of charge induced at the surface. In Sec.~\ref{sec:vib} we investigate the effect of the electric field on vibrational properties and compute the electron-phonon coupling constant and the superconductive critical temperature as a function of the induced charge density by using the Allen-Dynes/McMillan equation. In Sec.~\ref{sec:Eliashberg} the critical temperature is calculated more accurately by solving the isotropic linearized single-band Eliashberg equations. Finally, conclusions are given in Sec.~\ref{sec:conlusion}.\\

\section{\label{sec:methods}Model and Methods}
\begin{figure}
\centering
\subfloat[]
	{ \includegraphics[width=0.60\linewidth]{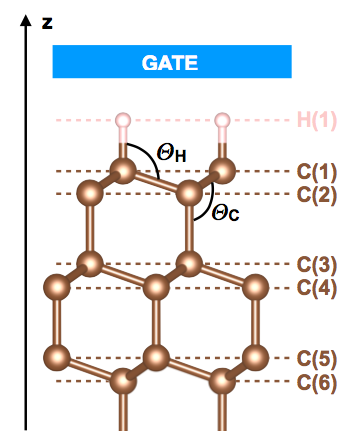}}
\subfloat[]
	{ \includegraphics[width=0.4\linewidth]{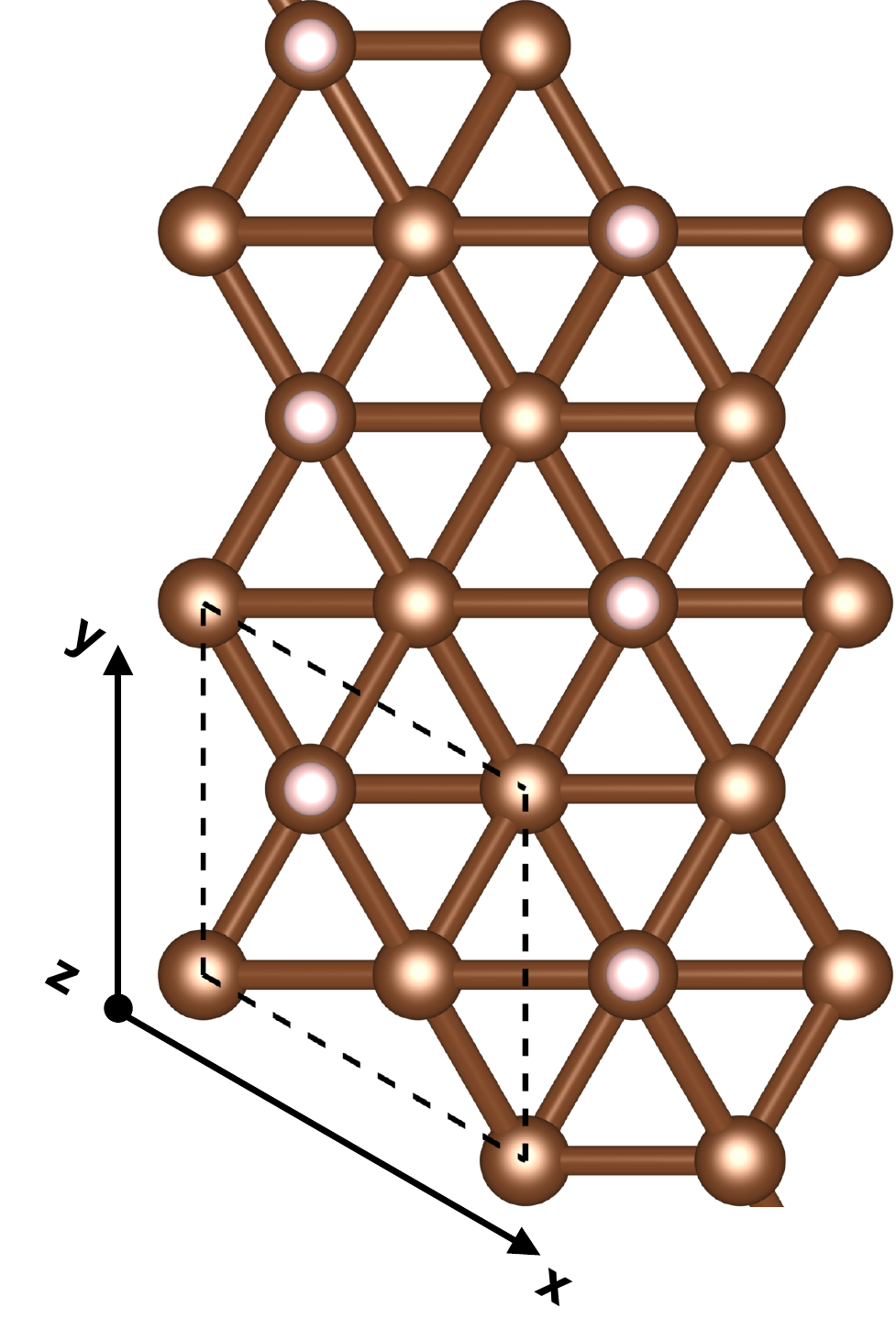}}
 \caption
   {
      Atomic structure of the hydrogenated (111) diamond surface:
      (a) side view,
      (b) top view.
      C atoms are the brown spheres while H atoms are the pink ones. Black dashed line identifies the unitary cell. $\Theta_H$ is the angle formed between H and C(2) atoms, while $\Theta_C$ is the angle between C(1) and C(3) atoms.
   }
   \label{fig:111_H}
\end{figure}

We consider a slab structure made of 14 layers of C atoms, oriented along the (111) direction, terminated on both sides by a layer of H atoms (fig.~\ref{fig:111_H}) for a total of 16 atoms per cell. The lattice parameter is taken coincident with that computed for bulk diamond. We analyze three different values of surface hole density, i.e. $n_{\text{dop},1}=2.84\cdot10^{13}$ cm$^{-2}$, $n_{\text{dop},2}=1.96\cdot10^{14}$ cm$^{-2}$ and $n_{\text{dop},3}=6.00\cdot10^{14}$ cm$^{-2}$. The first one corresponds to the experimental situation studied by Yamaguchi et al.~\cite{Yamaguchi}, who found the hydrogenated (111) diamond surface to be on the verge of an insulator-to-metal phase transition. The other two doping regimes are studied in order to see what would happen if higher values of the induced charge could be experimentally reached.

All the ab-initio calculations are performed using density functional theory with Quantum ESPRESSO package\cite{QE,QE_2,Thibault}. We use the Perdew-Burke-Ernzerhof (PBE) functional for exchange correlation, and the Brillouin zone integration is performed with a  $24\times24$ electron momenta (k-points) Monkhorst-Pack grid both for the neutral and charged systems. For hydrogen atoms we use ultrasoft pseudopotentials, while for carbon we use a norm-conserving pseudopotential.

The convergence criteria for the self-consistent solution of the Kohn-Sham equations are set to $10^{-9}$ Ry ($1\text{ Ry }\approx13.6\text{ eV}$) for the total energy and to $10^{-3}$ Ry/a$_0$ (a$_0\approx0.529177\text{ }\si{\angstrom}$ is the Bohr radius) for the maximum force acting on atoms. We also set the kinetic energy cut-off to $65$ Ry for the valence electron wave function and to $600$ Ry for the density.

The starting geometry for the FET configuration is similar to that described in Ref.\cite{Thibault}. Let $z$ be the axis perpendicular to the surface of the sample. We define a cell of length $L$ such that the slab is symmetric around $z=0$. The layer of accumulated ions at the surface is simulated through a sheet of uniformly distributed charges placed at $z_{gate}=-0.181L$. A potential barrier with a height of $V_0=3.5$ Ry is set at $z_{barrier}=-0.18L$ in order to avoid spilling of charges towards the metallic gate. Between two successive repeated images of the system we put $\approx30\text{ }\si{\angstrom}$ of vacuum. Ground state and linear response calculations are performed with the appropriate boundary conditions by truncating the Coulomb interaction in the non-periodic ($z$) direction.

We use Gaussian smearing of $0.004\text{ Ry}$ (for $n_{\text{dop},1}$), $0.03\text{ Ry}$ (for $n_{\text{dop},2}$) and $0.006\text{ Ry}$ (for $n_{\text{dop},3}$). Their value is chosen in such a way that the converged total energy per atom has a variation $<1$ mRy upon changing the value of the smearing. We also require that their magnitude is less then the energy difference between the top of the valence band and the Fermi level. When computing the electronic density of states (DOS) we used a tetrahedra smearing with a k-point grid of $48\times48$, which ensures better convergence of the results.

The convergence of phonon modes is checked at $\vb{q}=\vb{\Gamma}$. Convergence is found using $24\times24$ uniformly distributed k-points and widths of the Gaussian smearing of 0.003 Ry, 0.02 Ry and 0.004 Ry, for the three doping values respectively. Also for linear response calculations the smearing is always smaller than the energy difference between the top of the valence band and the Fermi level.

Electron-phonon computations are initially performed only at $\Gamma$ for all three doping regimes. For the highest surface charge densities, we also perform an interpolation of the electron-phonon matrix elements on the whole Brillouin zone through Wannier functions via the procedure described in Ref.~[16]. We can Wannier-interpolate the electronic band structure\cite{W90} using a grid of $6\times6$ k-points for the non-self consistent computation. We then interpolate the electron-phonon matrix elements to $75\times75$ k-points and phonon momenta (q-points) grids, which ensures convergence of quantities of interests.

\section{\label{sec:results}Results and Discussion}

An electric field applied at the surface of a material with free charge carriers penetrates only few layers inside the material, because of electronic screening. This kind of perturbation has different effects on our system: it breaks the symmetry along the $z$ direction (possibly lifting degeneracies in the electronic spectrum); it changes the charge distribution of electrons; and it modifies the relative positions of atoms (and therefore structural and vibrational properties).

First of all we analyze the effects of the applied electric field on the structure parameters of our sample. Then we study how the charge distribution, the electronic structure and the phonon dispersion are affected. Finally we tackle the problem of electron-phonon interactions and see if a superconductive phase can appear as a function of doping.

\subsection{\label{sec:relax}Structure Relaxation in FET Geometry}

The most relevant structural changes induced by the electric field occur in the first three layers of the slab, while the inner carbon atoms are barely affected.

In the lowest doping regime ($n_{\text{dop},1}$), the modifications are very subtle and only the distance between hydrogen atoms and the first carbon layer (H-C(1)) is shortened by $0.14\%$, while other atomic distances vary by $\sim 0.03- 0.05\%$. Both the angles $\widehat{HC(1)C(2)}$  and $\widehat{C(1)C(2)C(3)}$ (that we will call $\Theta_H$ and $\Theta_C$ in the following) remain unaffected (see Fig.~\ref{fig:111_H}).

As we move from low to medium values of the induced charge density ($n_{\text{dop},2}$), atomic distances further differ from equilibrium values. The H-C(1) bond is reduced  by $0.72\%$ while the distance C(1)-C(2) has an increment of $0.37\%$. As a consequence, also the angles $\Theta_H$ and $\Theta_C$ increase by $0.57\%$. The C(2)-C(3) bond is the least affected, with a reduction of only $0.25\%$.

Finally, at the highest doping ($n_{\text{dop},3}$) all three layers suffer considerable modifications. The H-C(1) bond is reduced by  $0.97\%$, the distance C(1)-C(2) is increased by $0.85\%$ and the C(2)-C(3) separation is shortened by $0.45\%$. The angles $\Theta_H$ and $\Theta_C$ are the most affected ones, with an increment of $1.26\%$.

From this analysis we can observe that variations of atomic distances get more substantial as we increase the amount of charges induced at the surface and, consequently, the magnitude of the electric field to which the sample is subjected. The structural modifications primarily concern the hydrogen atoms and the first layer of carbon atoms, as they directly face the metallic gate, but they also involve the second and third carbon layers with smaller intensities since the electric field is screened by the charge distribution inside the material.

The changes in the chemical bonds also suggest a redistribution of the electronic density inside the material due to the electrochemical gating, as it will be shown in the next subsection: indeed, as the magnitude of the electric field is increased, electrons tend to accumulate in certain spatial regions while depleting others, and this results in a variation of atomic distances. The results are summarized in Tab.~\ref{tab:111_struct_FET}.\\

\begin{table} [h]
\begin{center}
\begin{tabular}{c|c|c|c|c}
$n_{dop}$ (cm$^{-2}$) & $0$ & $2.84\cdot10^{13}$ & $1.96\cdot10^{14}$ & $6\cdot10^{14}$ \\
\hline
H - C(1) &  $1.109\si{\angstrom}$ &  $1.107\si{\angstrom}$ &  $1.101\si{\angstrom}$ &  $1.098\si{\angstrom}$\\
C(1) - C(2) &  $1.536\si{\angstrom}$ &  $1.537\si{\angstrom}$ &  $1.542\si{\angstrom}$ &  $1.549\si{\angstrom}$\\
C(2) - C(3) &  $1.554\si{\angstrom}$ &  $1.555\si{\angstrom}$ &  $1.550\si{\angstrom}$ &  $1.547\si{\angstrom}$\\
$\Theta_H$ & $108.55\si{\degree}$ & $108.64\si{\degree}$ & $109.18\si{\degree}$ & $109.92\si{\degree}$ \\
$\Theta_C$ & $108.55\si{\degree}$ & $108.64\si{\degree}$ & $109.18\si{\degree}$ & $109.92\si{\degree}$ \\
\hline
\end{tabular}
\caption{Structural parameters of the hydrogenated (111) diamond surface in FET geometry at different doping levels. For angles and atom labeling see Fig.~\ref{fig:111_H}}
\label{tab:111_struct_FET}
\end{center}
\end{table}

\subsection{\label{sec:exfoliation}FET-driven Exfoliation of Surface States}

\begin{figure}[h]
\centering
\includegraphics[width=1.0\linewidth]{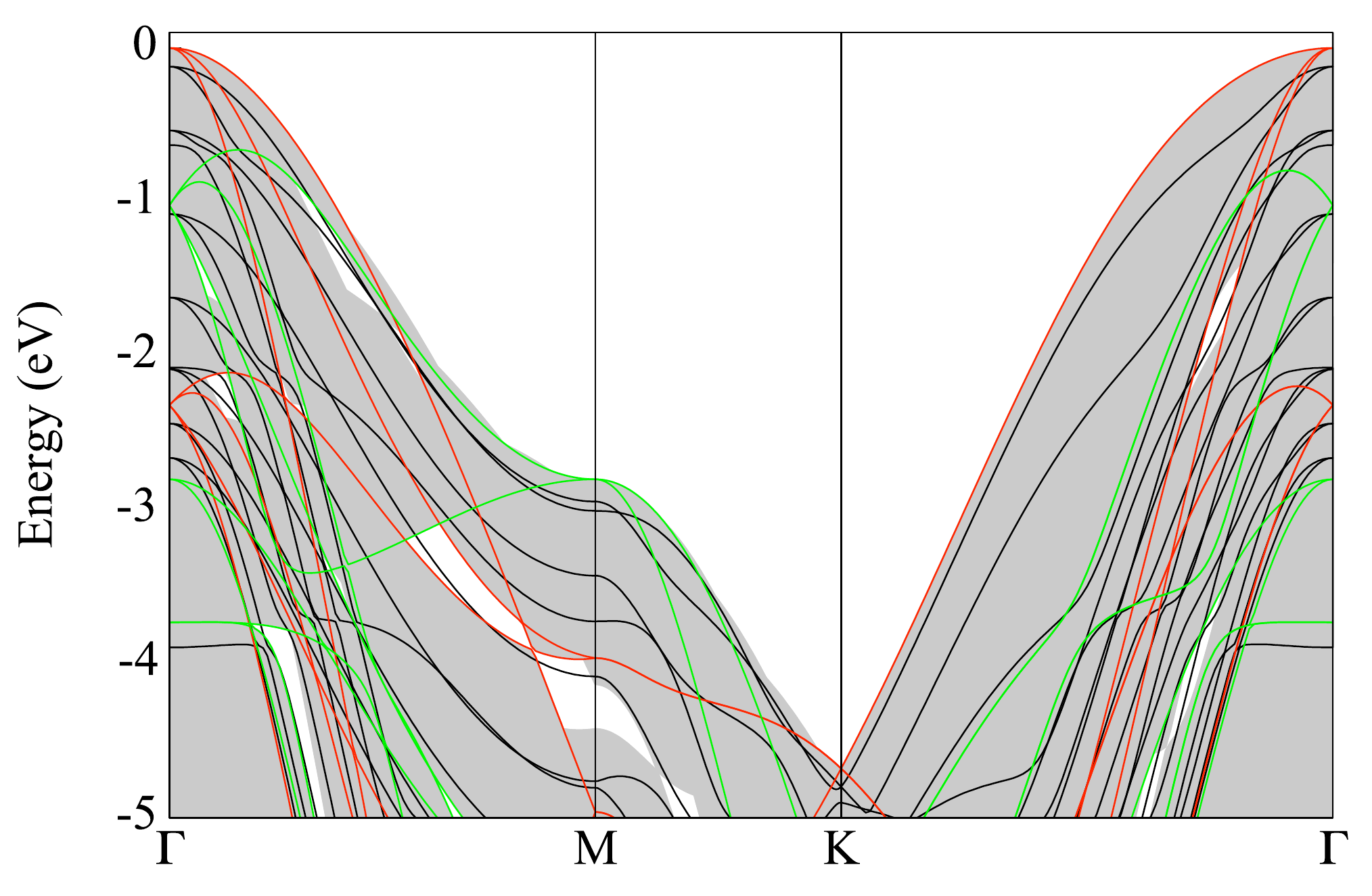}
 \caption{Band structure of the undoped hydrogenated (111)-diamond surface (black lines) on top of the surface-projected bulk-diamond electronic structure (grey shaded regions). Red (green) lines correspond to the bulk band structure computed for $k_z=0.0$ ($k_z=0.5$) in crystal coordinates.}
   \label{fig:BO_U}
\end{figure}

\begin{figure}
\centering
\subfloat[]
	{ \includegraphics[width=1.0\linewidth]{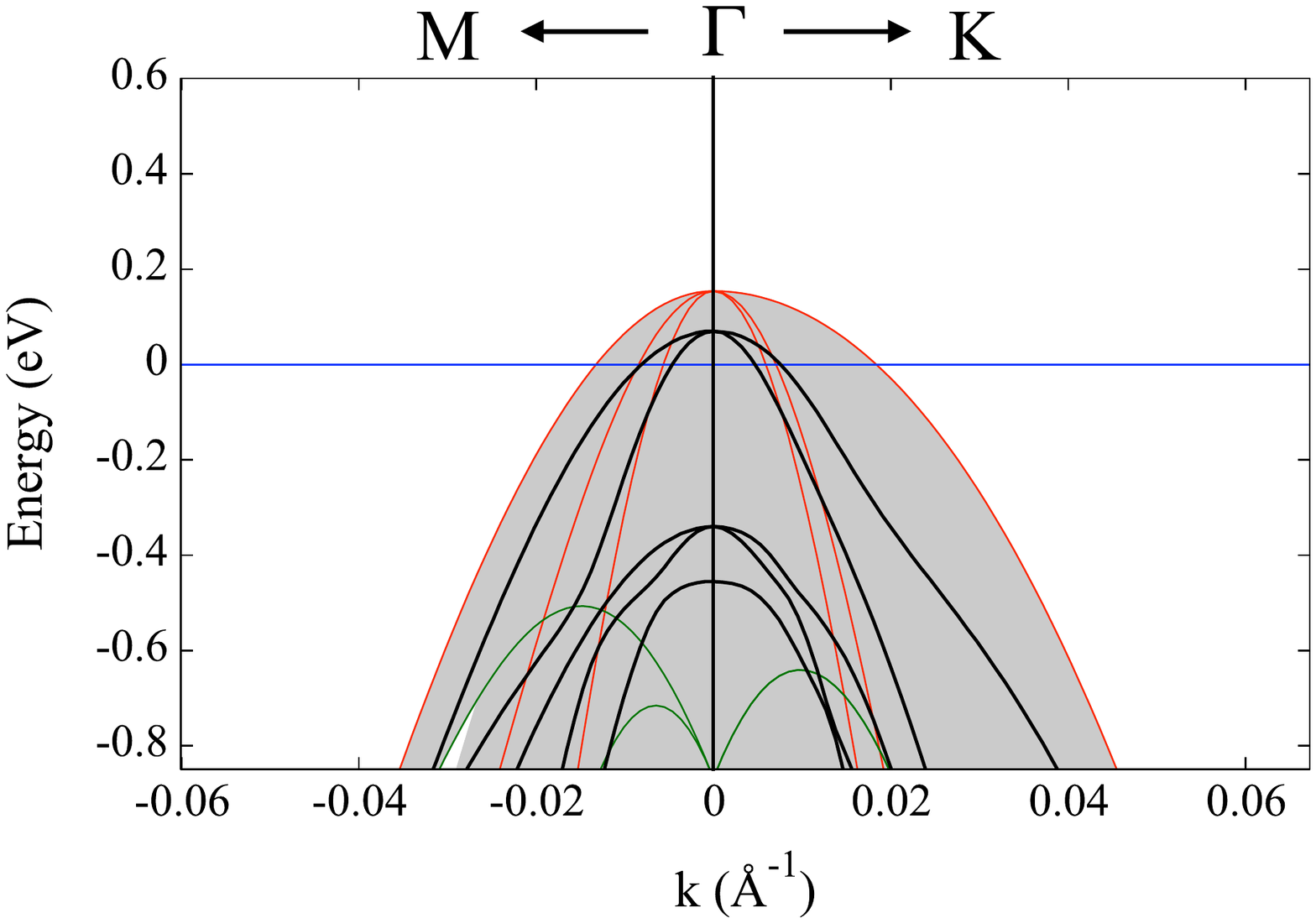}}\\
\subfloat[]
	{ \includegraphics[width=1.0\linewidth]{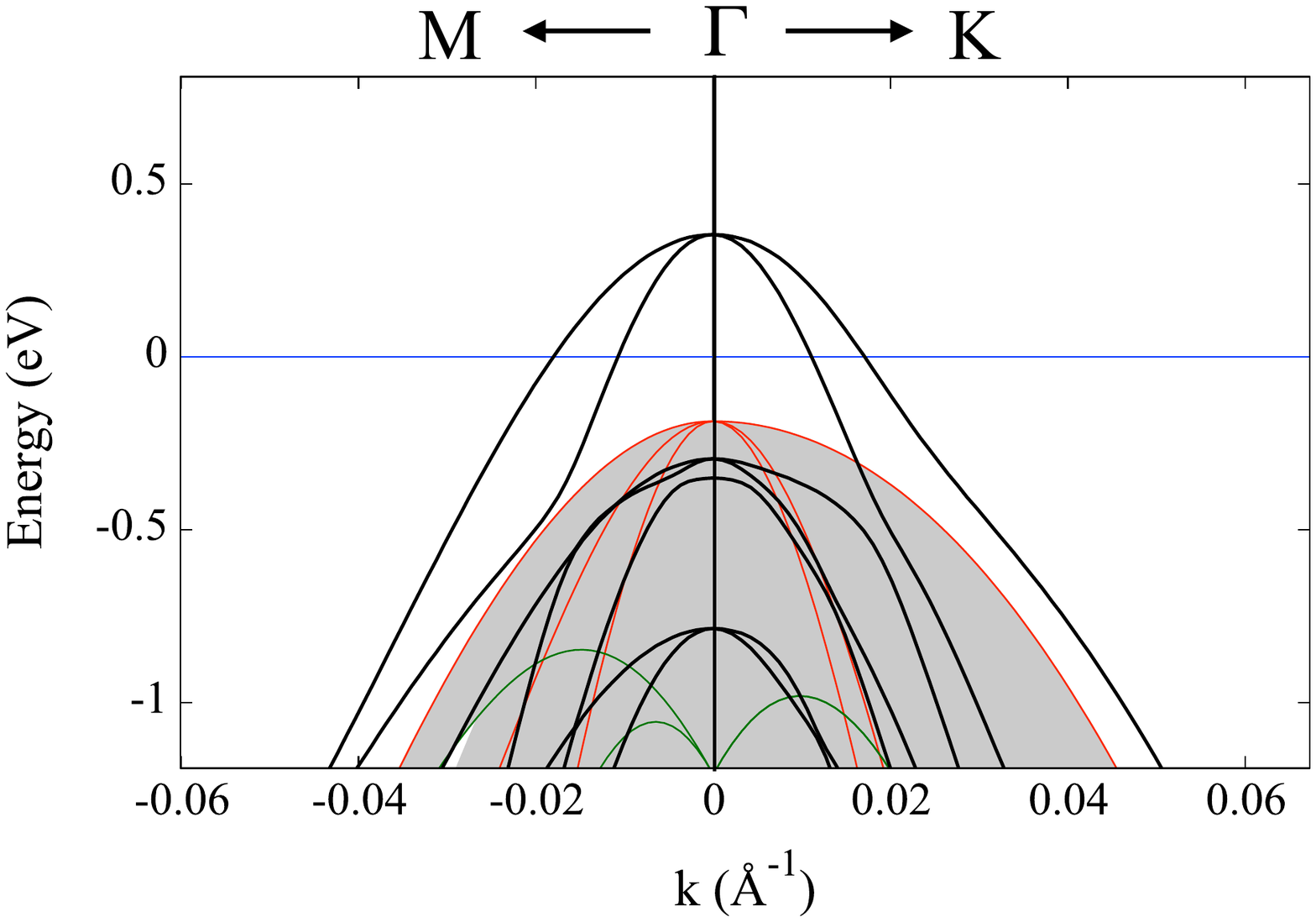}}\\
\subfloat[]
	{ \includegraphics[width=1.0\linewidth]{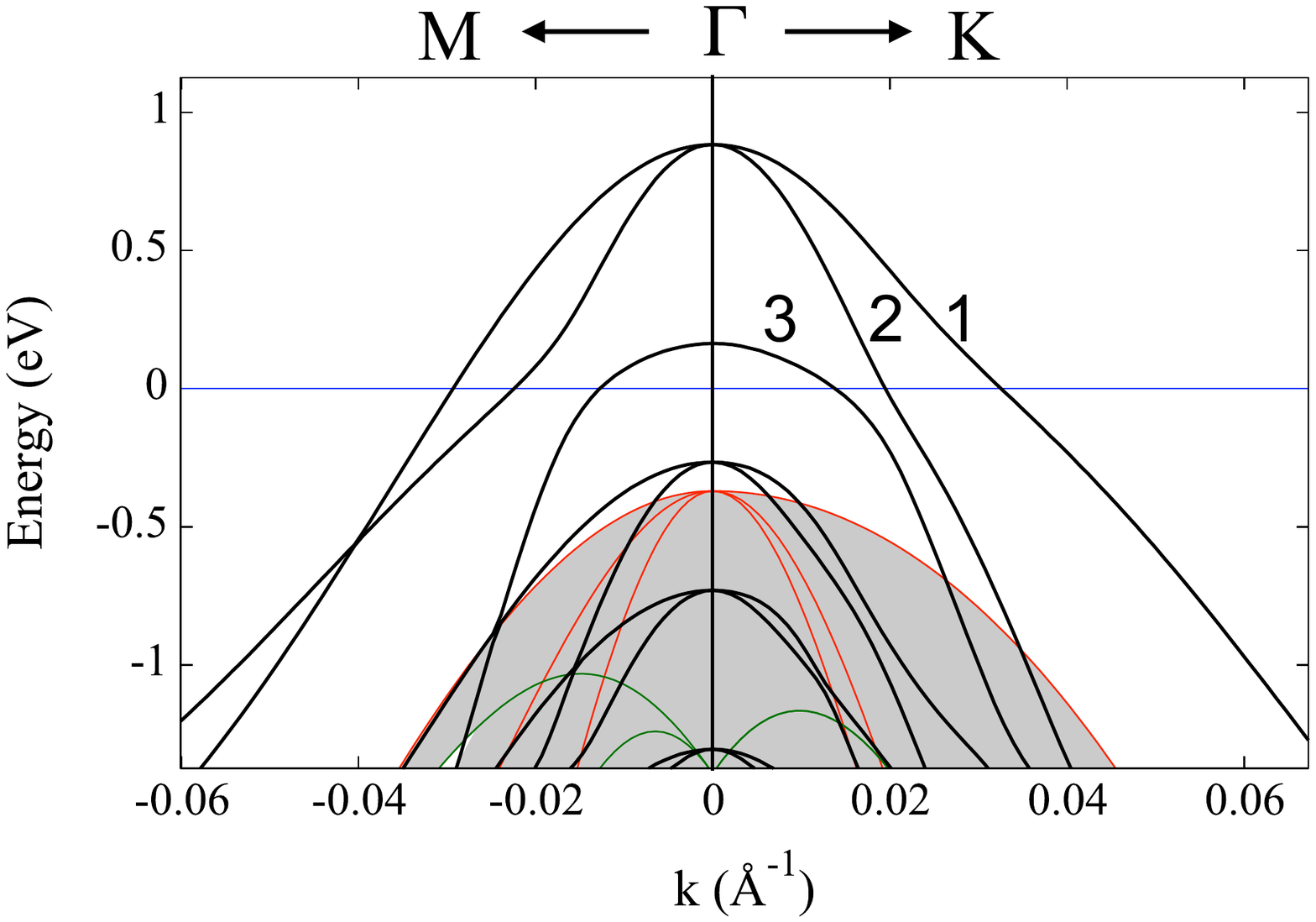}}\\	
 \caption
   {Band structure of the doped hydrogenated (111)-diamond surface (black lines) on top of the surface-projected bulk-diamond electronic structure (grey shaded regions). The three doping regimes are: (a) $n_{\text{dop},1}$, (b) $n_{\text{dop},2}$ and (c) $n_{\text{dop},3}$.  Red (green) lines correspond to the bulk band structure computed for $k_z=0.0$ ($k_z=0.5$) in crystal coordinates. Blue line is the Fermi level. The bands have been aligned as described in Sec.~\ref{sec:exfoliation}.
   }
   \label{fig:BO_D}
\end{figure}

In order to understand which are the surface and bulk electronic states, we plot the band structure of the hydrogenated (111)-diamond slab on top of the surface-projected bulk-diamond electronic bands.
The latter quantity has been obtained by taking diamond in its bulk form oriented along the (111) direction and then computing its band diagram by keeping fixed $\vb{k}_{||}=$($k_x$,$k_y$) along paths parallel to $\Gamma$-M-K-$\Gamma$ while varying $k_z$ between $0$ and $\pi/c$ (where $c$ is the height of the primitive cell). This procedure generates a continuum of bulk bands that, if plotted altogether in the same graph, give rise to the gray-shaded regions in Fig.~\ref{fig:BO_U} and Fig.~\ref{fig:BO_D}.
When the band structure of the slab is superimposed to this graph, the bands that fall outside the grey regions can be identified as surface states, while the others are considered as bulk states.

However, the two band diagrams (bulk and slab) have to be properly aligned in energy before being superimposed to each other. As a matter of fact the surface, that is the interface between vacuum and the material under investigation, perturbs the electronic charge distribution and generates an electrostatic potential discontinuity. This is the same problem as the band alignment between two different semiconductors or at a metal/semiconductor interface: indeed in these cases a Schottky barrier is formed, whose height measures the valence band discontinuity across the interface. By lining up the electrostatic potential on the two sides of the surface we can compute the band offset, that is the difference between the top of the valence band of the surface and that of the remaining bulk of the material.

We define the planar-averaged electrostatic potential as:

\begin{equation}
V_{||}(z)= \frac{1}{\Omega_{2D}}\int_{\Omega_{2D}}V_{3D}(x,y,z)dxdy
\end{equation}
where $V_{3D}(x,y,z)$ is the 3D electrostatic potential and $\Omega_{2D}$ is the area of the primitive cell. Then, in order to extract only the relevant interface-related phenomena, we compute the \emph{macroscopic average}\cite{Baldareschi} of the planar-averaged electrostatic potential:

\begin{equation}
\tilde{V}_{||}(z) = \frac{1}{a}\int_{z-a/2}^{z+a/2}V_{||}(s)ds
\end{equation}
that is an average of the local electrostatic potential performed inside a window of width $a$, whose size depends on the system under study and has to be big enough in order to filter-out the microscopic details of the material.

The computational procedure goes as follows:

\begin{itemize}
\item We take the hydrogenated (111)-diamond and compute the macroscopic average of the planar-averaged electrostatic potential. Its value in the bulk-like region of the slab ($\epsilon_{\text{RS}}$) will serve as the reference energy for the surface system. Then we define $\Delta\epsilon^{\text{S}}=\epsilon_{\text{VB}}^{\text{S}}-\epsilon_{\text{RS}}$ as the difference between the energy of the top of the valence band ($\epsilon_{\text{VB}}^{\text{S}}$) and the reference energy of the slab;\\
\item We take diamond in its bulk form and compute the macroscopic average of the planar-averaged electrostatic potential. Its mean value ($\epsilon_{\text{RB}}$) will serve as the reference energy for the bulk system. Then we define $\Delta\epsilon^{\text{B}}=\epsilon_{\text{VB}}^{\text{B}}-\epsilon_{\text{RB}}$ as the difference between the energy of the top of the valence band ($\epsilon_{\text{VB}}^{\text{B}}$) and the reference energy of the bulk;\\
\item The energy shifts between the band structure of the slab and that of the bulk projected over the surface is given by $\Delta\epsilon^{\text{SB}}=\Delta\epsilon^{\text{S}}-\Delta\epsilon^{\text{B}}$. This is the quantity by which the surface bands have to be shifted before being plotted on top of the bulk projected bands.
\end{itemize}

This procedure has to be applied both for the undoped and doped surface.
It turns out that, when the surface is undoped (Fig.~\ref{fig:BO_U}) the electronic bands are all bulk states since they are inside the grey-shaded areas. Upon doping the band shift increases, but for low values of the induced charge ($n_{\text{dop},1}$) we still are in a bulk-like situation (Fig.~\ref{fig:BO_D}(a)). As we will see in the next section (Fig.~\ref{fig:111_charge}(a)) the charge spreads well inside the slab, meaning that the surface is still electronically indistinguishable from the bulk. However, when we pass to medium and high doping values ($n_{\text{dop},2}$ and $n_{\text{dop},3}$) the bands crossing the Fermi level acquire a surface-like character, as clearly shown in Fig.~\ref{fig:BO_D}(b) and Fig.~\ref{fig:BO_D}(c). We will refer to this phenomenon as \emph{electric exfoliation}. This will be confirmed by our subsequent analysis of the planar-averaged charge density (Fig.~\ref{fig:111_charge}(b) and Fig.~\ref{fig:111_charge}(c)).

\subsection{\label{sec:chrg}Charge Distribution}
\label{sec:charge}

\begin{figure*}
\centering
\subfloat[]
	{ \includegraphics[width=0.45\linewidth]{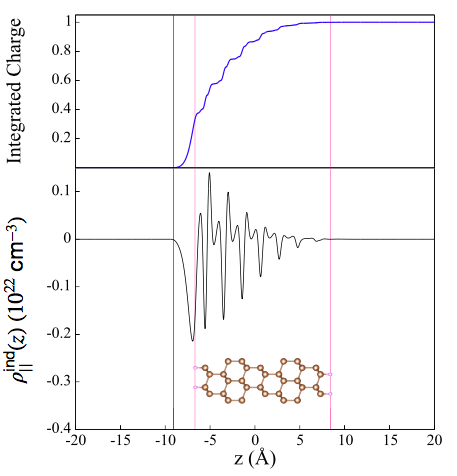}}\quad\quad
\subfloat[]
	{ \includegraphics[width=0.47\linewidth]{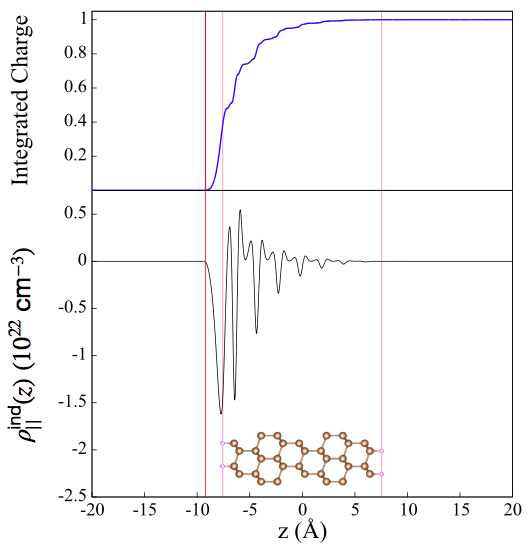}}\\
\subfloat[]
	{ \includegraphics[width=0.47\linewidth]{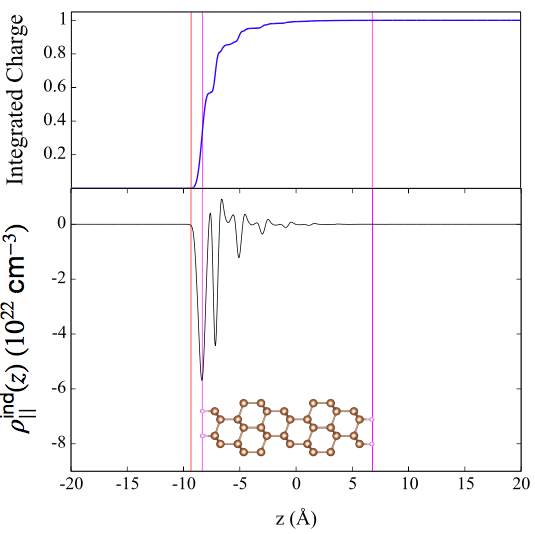}}
   \caption
 {Planar-averaged induced charge density ($\rho^{\text{ind}}_{||}(z)$) and normalized cumulative integral of the absolute value of the planar-averaged charge density (integrated charge). The three cases correspond to (a) $n_{\text{dop},1}$, (b) $n_{\text{dop},2}$ and (c) $n_{\text{dop},3}$. The red lines indicates the location of the potential barrier which prevents charge spilling, while the pink lines identify the position of the hydrogen layer which also show the beginning and the end of the diamond surface.
   }
   \label{fig:111_charge}
\end{figure*}

In order to understand how many layers are affected by the electrochemical gating, it is important to see how the charge density profile evolves as we go from a low to a high doping regime.

In the lower panels of fig.~\ref{fig:111_charge} (a), (b) and (c) we display the planar-averaged induced charge density for all the three values of doping:

\begin{equation}
\rho^{\text{ind}}_{||}(z)=\rho^{\text{h}}_{||}(z)-\rho^{0}_{||}(z)
\end{equation}

Here $\rho^{\text{h}}_{||}(z)$ and $\rho^{0}_{||}(z)$ are respectively the planar-averaged charge densities for the hole-doped and undoped cases:

\begin{equation}
\rho^{\text{i}}_{||}(z)= \frac{1}{\Omega_{2D}}\int_{\Omega_{2D}}\rho^{\text{i}}_{3D}(x,y,z)dxdy
\end{equation}
where $\Omega_{2D}$ is the area of the unitary cell and $\rho^{\text{i}}_{3D}(x,y,z)$ is the 3D charge density for the hole-doped ($\text{i}=\text{h}$) and undoped ($\text{i}=0$) cases. Both the undoped and doped charge densities are computed from the slab with atomic parameters obtained from the relaxation in presence of the electric field.\\
The 3D charge density is given by:

\begin{equation}
 \rho^{\text{i}}_{3D}(x,y,z) = \frac{e}{N_{k}}\sum_{\sigma,\vb{k}\in\text{FBZ}}\sum_{n=0}^{E_F}\abs{\psi_{\vb{k},n}^{\sigma}(x,y,z)}^2\, .
\end{equation}

Here $e$ is the electron charge, $N_{k}$ is the number of k points belonging to the first Brillouin zone (FBZ), $\sigma$ is the spin index, $n$ is the band index, $E_F$ is the Fermi energy and $\psi^{\sigma}_{\vb{k},n}(x,y,z)$ is the Bloch wavefunction for band $n$, spin $\sigma$ and k-point $\vb{k}$.

These charge profiles contain information on both the amount of carriers induced in the sample and on how pre-existing valence electrons re-arrange themselves in order to screen the applied electric field~\cite{Brumme}. Regions in which holes are accumulated (i.e. electrons are depleted) correspond to negative values of $\rho^{\text{i}}_{||}$, while regions enriched with electrons are characterized by positive values of  $\rho^{\text{i}}_{||}$. In the low doping regime ($n_{\text{dop},1}$) charge fluctuations are considerable up to the sixth carbon layer, which indicates that we are still in a bulk-like situation with a poor localization of charges at the surface. As we induce more charges at the surface ($n_{\text{dop},2}$), fluctuations become more peaked around the first four carbon layers and, eventually, at the highest doping regime ($n_{\text{dop},3}$) they are strongly localized in proximity of the first two carbon layers, meaning that we are moving to a quasi-2D system as the screening effect of charges on the electric field becomes stronger.

We can also get information on where charges are localized in these three cases by plotting the normalized cumulative integral of the absolute value of the planar-averaged charge density:

\begin{equation}
\frac{1}{\int_{0}^L\abs{\rho^{\text{i}}_{||}(z)}dz}\cdot\int_{0}^c\abs{\rho^{\text{i}}_{||}(z)}dz\quad\text{ for }c\in[0;L]
\end{equation}
where $L$ is the size of the cell. This quantity is shown in the upper panels of fig.~\ref{fig:111_charge} (a), (b) and (c): at the beginning charges are more spread inside the system, as $70\%$ of the integrated charge is reached after three atomic layers (i.e. 6 carbon layers) but, as we increase the doping, this limit is attained in close proximity to the surface, i.e. in the region containing the first two carbon layers.

\subsection{\label{sec:elec}Electronic Properties}

\begin{figure}
\centering
\subfloat[]
	{ \includegraphics[width=\linewidth]{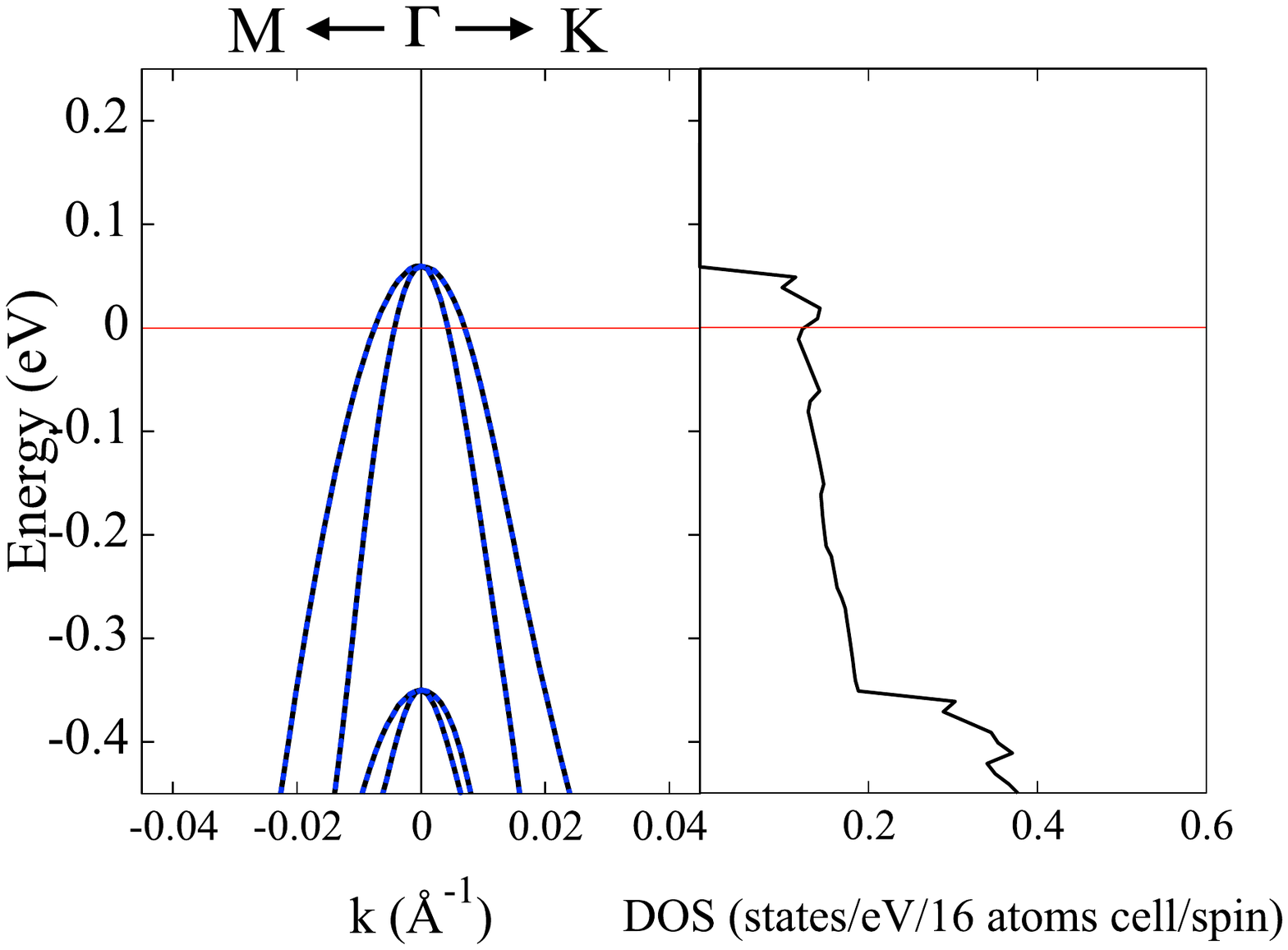}}\\
\subfloat[]
	{ \includegraphics[width=\linewidth]{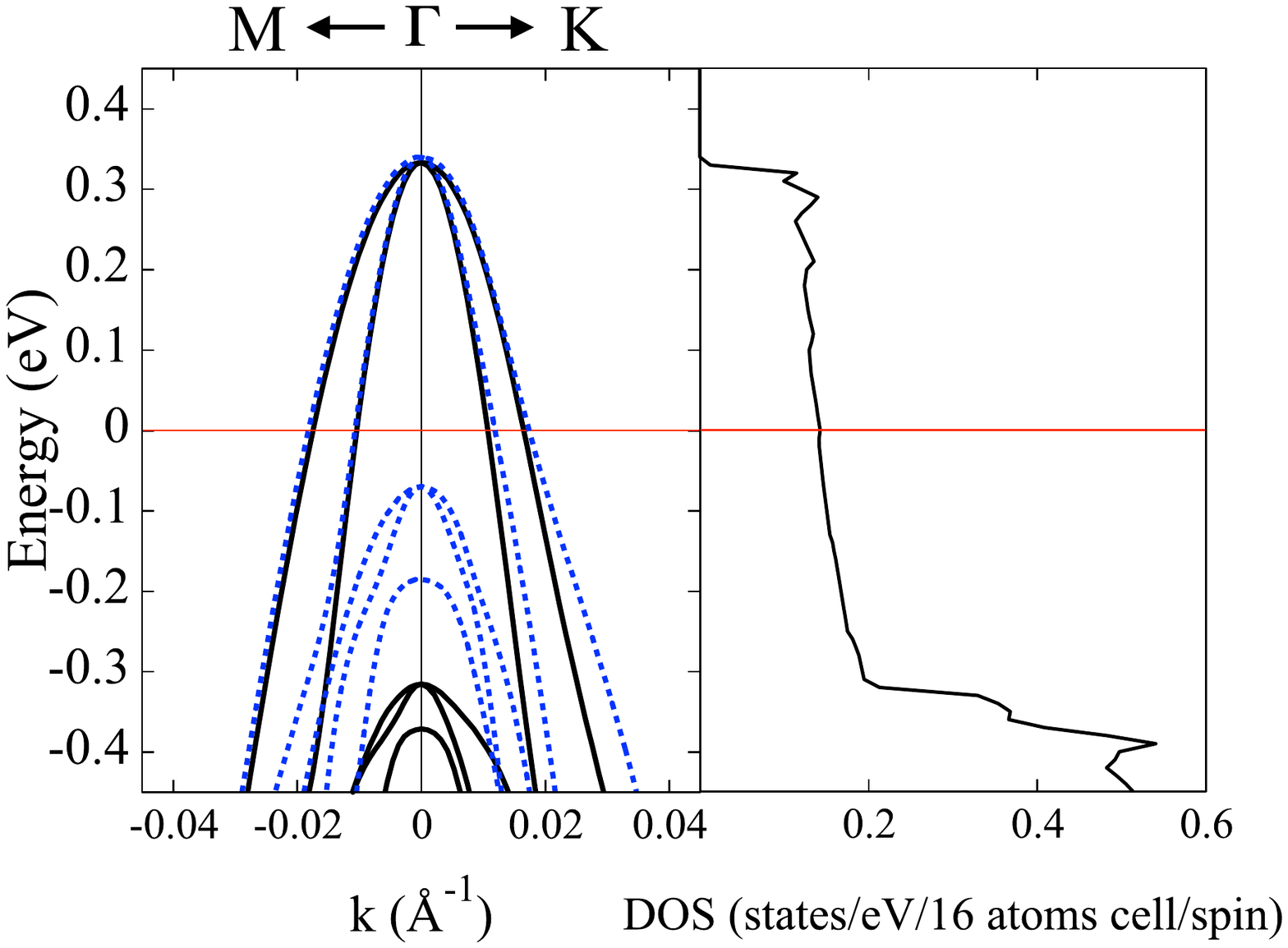}}\\
\subfloat[]
	{ \includegraphics[width=\linewidth]{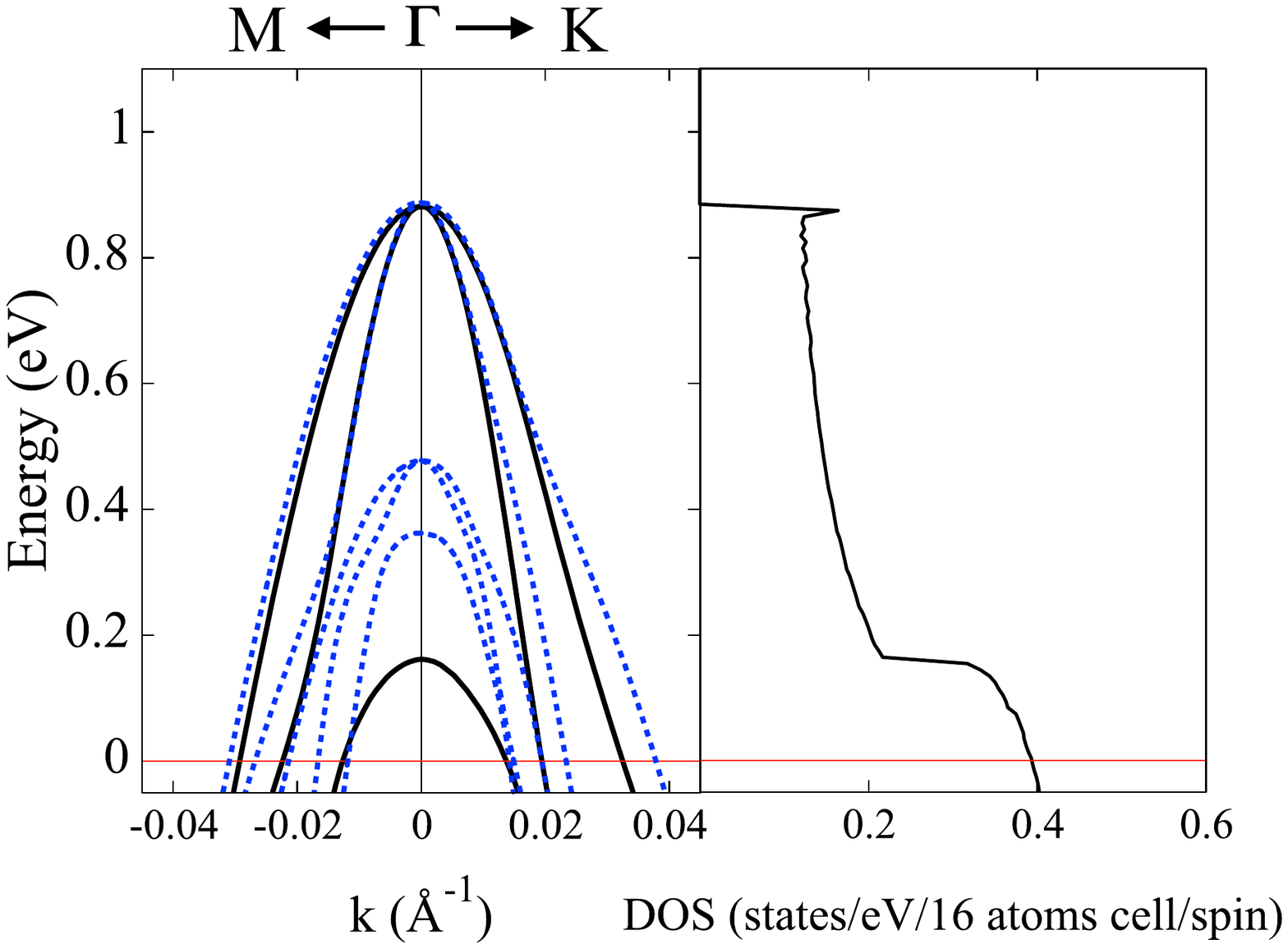}}\\	
 \caption
   {Electronic band structure and density of states (DOS) of the hydrogenated (111) diamond surface with (a) $n_{\text{dop},1}$, (b) $n_{\text{dop},2}$ and (c) $n_{\text{dop},3}$. As a reference (blue dashed lines) we also plot the uniform doping as obtained from jellium. Black solid lines refer to the FET case, the red line is the Fermi Energy.
   }
   \label{fig:111_bands}
\end{figure}

After the analysis of the charge distribution we can study how electronic properties are affected by the electrochemical gating. In Fig.~\ref{fig:111_bands} we analyze the electronic band structure and density of states of the diamond surface as a function of the induced charge density. In this figure we compare the bands computed in FET geometry (black solid line) with the ones obtained by using a jellium model (blue dashed line), i.e. a model where doping is modeled with charges uniformly spread inside the slab and compensated by an oppositely charged background.

The Fermi level, $E_F$ (red line), appears crossing the valence bands: at the lowest charge doping it is right below the top of the first two valence bands (bands will be referred to according to the labelling of Fig.~\ref{fig:BO_D} (c)) but, as the induced charge density gets higher, $E_F$ gets lower and lower and it even crosses a third band. Moreover the presence of the electric field breaks the symmetry along the $z$ direction, lifting any band degeneracy. It is possible to observe this effect on the third crossed band (see Fig.~\ref{fig:111_bands}(c)), as its double degeneracy at the center of the FBZ is lifted isolating the only band which is crossed at the highest doping. This suggests the nature of the bands, i.e. a planar ([xy]) character for the first two bands and an out-of-plane ([z]) behavior for the third band. Finally, we can also observe that the energy gap ($E_G$) is also affected by the presence of the electric field: valence bands are shifted upward in energy while conduction bands are shifted downward gradually reducing the band gap as we increase the magnitude of the electric field. Therefore it is clear that the jellium model is not suitable to describe the FET geometry: in fact, it would have led to a simple rigid shift of the occupied energy bands preserving all possible degeneracies of the eigenvalues. These results are summarized in Tab.~\ref{tab:111_energies}.\\

\begin{table}
\begin{center}
\begin{tabular}{c|c|c|c|c}
$n_{dop} $ (cm$^{-2}$)& $E_G$ & $\Delta E_{V,I}$& $\Delta E_{V,II}$ & $\Delta E_{V,III}$ \\
\hline
$2.84\cdot10^{13}$ & 3.516 eV & 58.6 meV&  58.6 meV & Not Crossed\\
$1.96\cdot10^{14}$ & 2.525 eV & 331.7 meV&  331.7 meV & Not Crossed\\
$6\cdot10^{14}$ & 1.717 eV & 880.2 meV & 880.2 meV& 161.6 meV\\
\hline
\end{tabular}
\caption{Energy gaps and positions of Fermi level with respect to the top of valence bands, $\Delta E_{V,i}$ $i=I,III$ of the hydrogenated (111) surface in FET geometry.}
\label{tab:111_energies}
\end{center}
\end{table}

The DOS shows the typical energy dependency of 2D systems (i.e. $\text{DOS}(\epsilon)\sim\text{const.}$). For the first doping value we already have a relatively high DOS at the Fermi level, due to two bands crossing, but it nearly triples when the third band is crossed as well (Tab.~\ref{tab:BDOS_111}). Since this is a multiband system, we are also interested in the contribution to the total density of states from each single energy band. This information will be important in the next subsection for understanding the origin of the electron-phonon interactions, since they are directly proportional to the density of states per spin per band. Band 1 contributes to the total DOS $\sim2-3$ times more than band 2 for all three doping values, while the contributions of bands 2 and 3 have a similar value (Tab.~\ref{tab:BDOS_111}). This is a preliminary clue that the most important electron-phonon interactions will take place in the first band.

\begin{figure}
\centering
\subfloat[]
	{ \includegraphics[width=\linewidth]{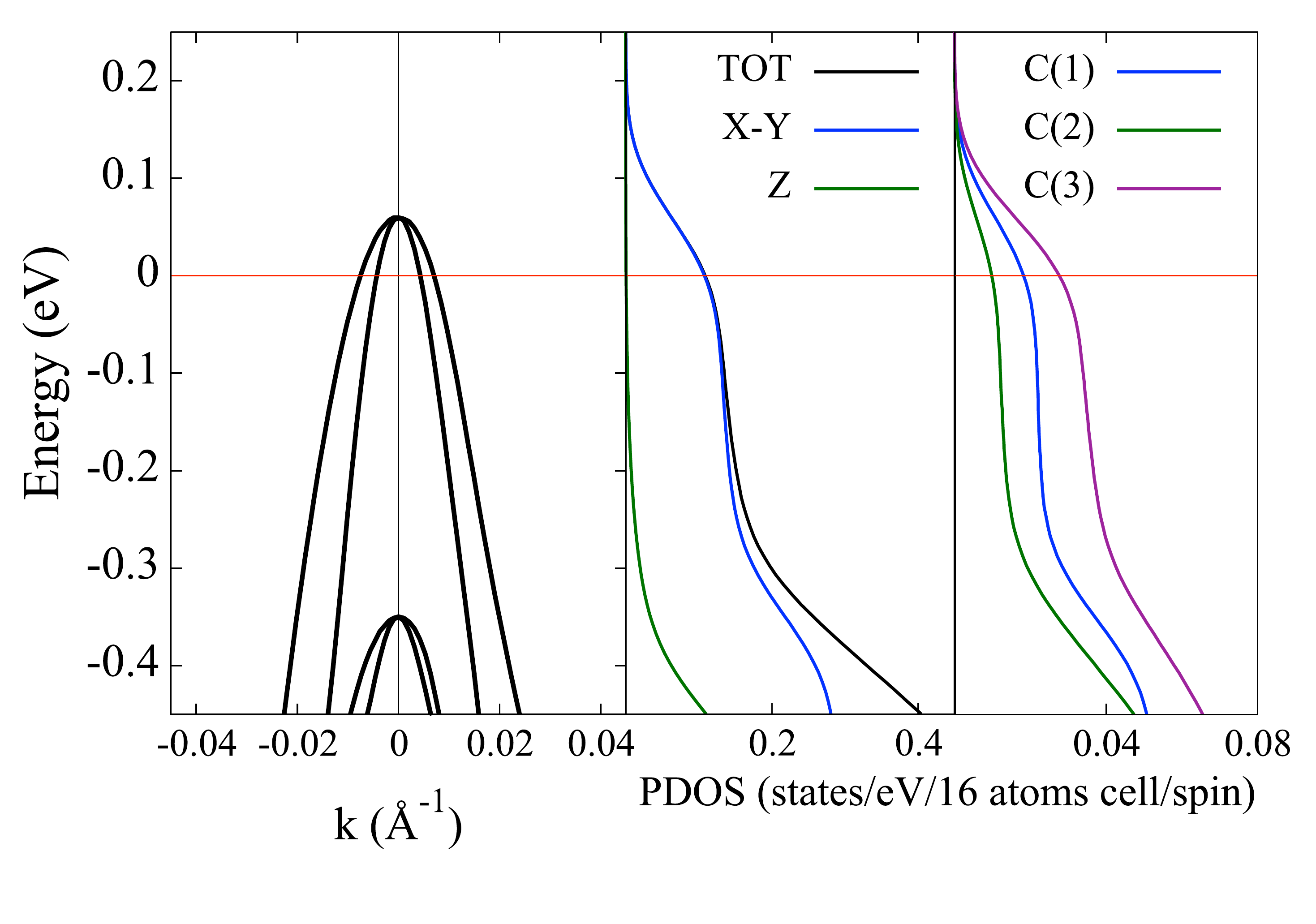}}\\
\subfloat[]
	{ \includegraphics[width=\linewidth]{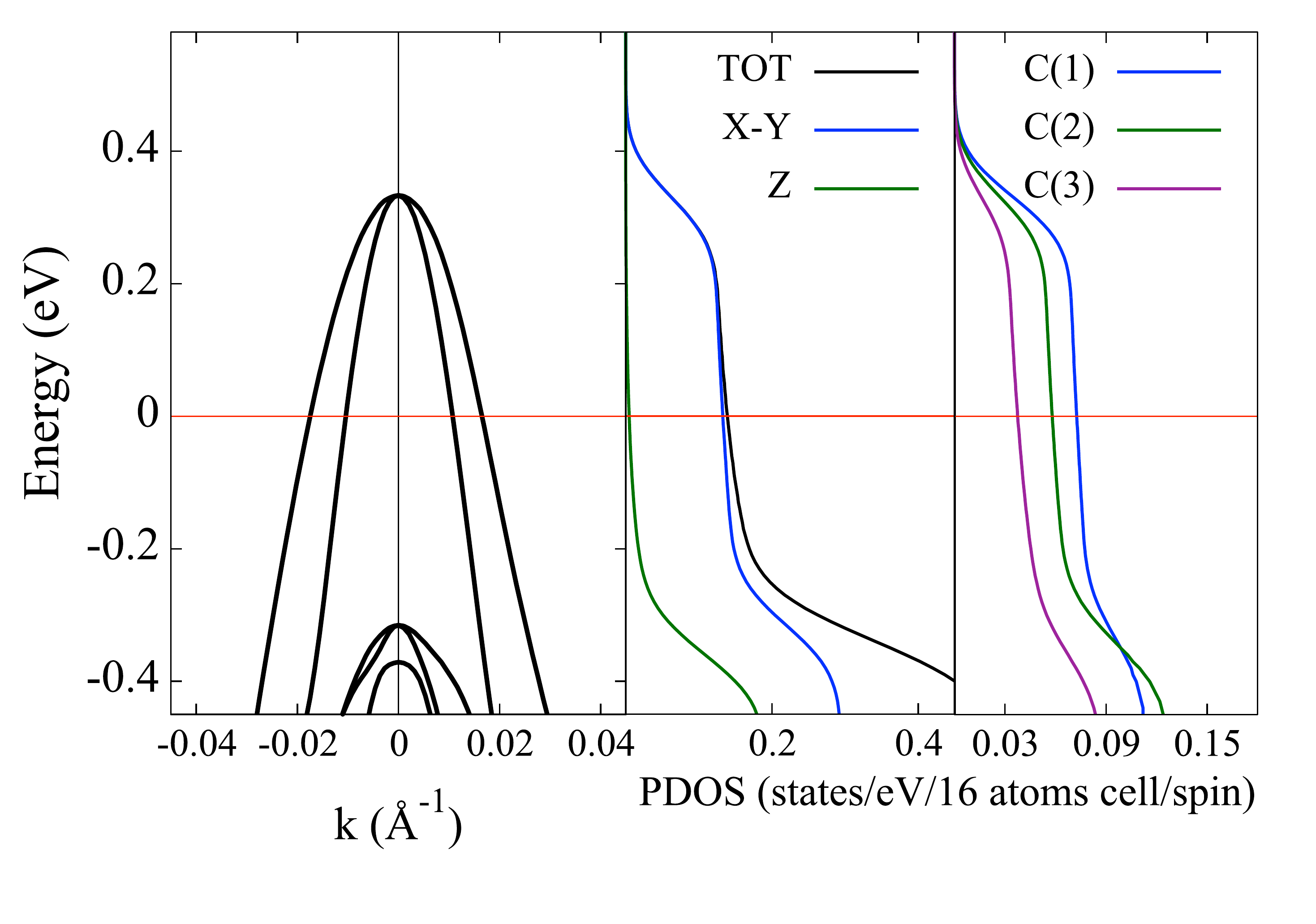}}\\
\subfloat[]
	{ \includegraphics[width=\linewidth]{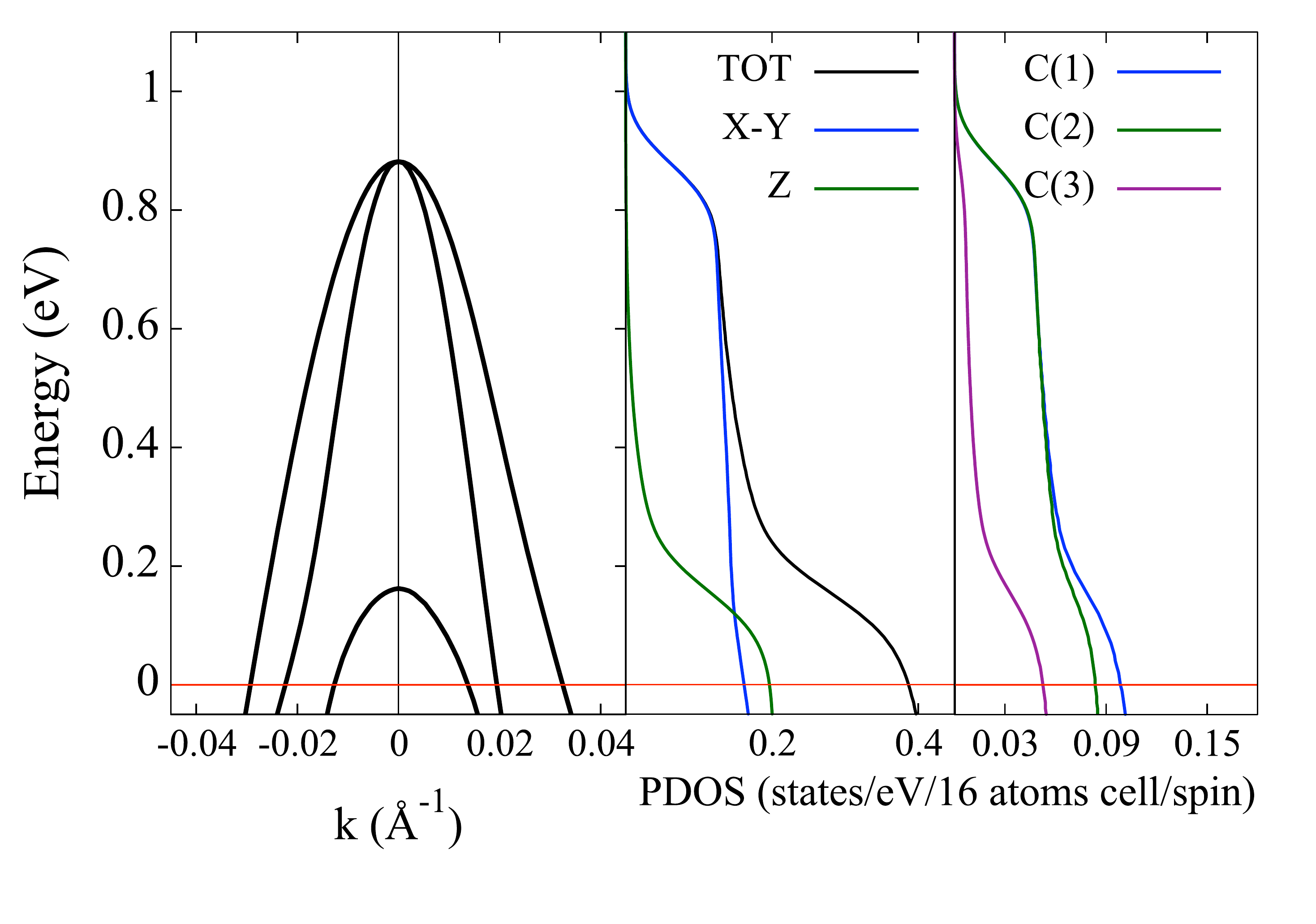}}\\	
 \caption
   {Electronic band structure and projected density of states (PDOS) of the hydrogenated (111) diamond surface with (a) $n_{\text{dop},1}$, (b) $n_{\text{dop},2}$ and (c) $n_{\text{dop},3}$. The red line is the Fermi Energy. Central panel gives the PDOS along in-plane ([xy]) and out-of-plane ([z]) components. Right panel shows the C(1), C(2) and C(3) actual contributions to the DOS.
   }
   \label{fig:111_pdos}
\end{figure}

The projected density of states on atomic orbitals (PDOS) can give us more details on the nature of the electronic bands. The central panels of Fig.~\ref{fig:111_pdos}, show the density of states projected over the in-plane ([xy]) and out-of-plane ([z]) components, while the right panels display the actual contribution to the DOS of the first three carbon layers (C(1), C(2) and C(3)) . First of all the PDOS confirms that bands 1 and 2 have an in-plane behavior while the out-of-plane contribution is present only when the band 3 is crossed. Another interesting aspect is that, at the Fermi level, the value of the in-plane PDOS is always higher than the out-of-plane one for low to medium doping values, while they become comparable at the highest doping: this will be important in the subsequent analysis of the vibrational properties in order to understand the nature of vibrational modes and of the electron-phonon interaction. Finally we can observe that the contribution of the C atom layers to the density of states are in agreement with the evolution of the charge density profile: indeed at the beginning the charge density oscillations are more delocalized inside the material and, as a consequence, the C(3) component is higher. When the induced charge density gets more and more confined towards the sample surface the C(1) and C(2) contributions to the DOS show a higher value at the Fermi level. The disappearance of the C(3) character from the valence band top is another effect of the electric exfoliation.

\begin{table}[h]
\begin{center}
\begin{tabular}{c|c|c|c|c}
$n_{dop} $ (cm$^{-2}$)&$N_{\text{tot}} (0)$ &$N_1$(0) & $N_2$(0) & $N_3$(0) \\
\hline
$2.84\cdot10^{13}$ &$0.1319 $ &$  0.0956 $ & $  0.0363 $ &   None \\
$1.96\cdot10^{14}$ & $0.1427 $&$ 0.1052 $ & $ 0.0375 $ & None \\
$6\cdot10^{14}$ & $0.3930$&$ 0.1783 $ & $ 0.1088 $ & $ 0.1059$\\
\hline
\end{tabular}
\caption{Total density of states $N_{\text{tot}} (0)$ and band contributions to the total density of states $N_{\text{i}} (0)$ ($\text{i}=1,3$) at the Fermi level ($E_F=0$) for the hydrogenated (111) diamond surface, in units of states/eV/16 atoms cell/spin. Bands are labeled according to Fig.~\ref{fig:BO_D} (c).}
\label{tab:BDOS_111}
\end{center}
\end{table}

The FET doping also induces a change in the curvature of the bands themselves that, obviously, indicates a change in the effective masses of the carriers (as shown in Fig.~\ref{fig:111_bands}).

If we approximate our bands by simple parabolas:

\begin{equation}
E(\vb{k}) = \hbar^2\Biggl(\frac{k_x^2}{2m_x^*} + \frac{k_y^2}{2m_y^*}\Biggr)
\end{equation}
the effective masses are proportional to the inverse of the band structure curvature:

\begin{equation}
\frac{1}{m^*_{ij}}=\frac{1}{\hbar^2}\pdv{E(\vb{k})}{k_i}{k_j}
\end{equation}

From Tab.~\ref{tab:eff_mass_111} we can see that the effective mass of band 1 decreases along  $\Gamma$-M and increases along $\Gamma$-K on going from zero doping to $n_{\text{dop},1}$, and then remains stationary. The effective mass of band 2 first decreases (in both directions), but also in this case it doesn't change anymore after the intermediate doping value. Finally, the effective mass of band 3 decreases along $\Gamma$-M and increases along $\Gamma$-K.

As for the band anisotropy, band 1 is initially quite anisotropic but recovers a parabolic symmetry when we increment the amount of induced charge density. Band 2 is symmetric already at the lowest doping, and does not change its character throughout the various doping regimes. Finally, band 3 is the one which has the highest anisotropy and this is accentuated as $n_{\text{dop}}$ increases.

\begin{table}
\begin{center}
\begin{tabular}{c|c|c|c}
&& $\abs{m^*_{\Gamma-M}/m_e}$& $\abs{m^*_{\Gamma-K}/m_e}$ \\
\hline
&Undoped& $0.186 $& $0.116 $  \\
&$n_{\text{dop},1}$ & $0.145 $& $0.117 $  \\
$\text{I}_{\text{band}}$&$n_{\text{dop},2}$ & $0.132 $& $0.126$  \\
& $n_{\text{dop},3}$ & $0.132 $& $0.126 $  \\
&&\\ \hline
&Undoped& $0.057 $& $0.059 $  \\
&$n_{\text{dop},1}$ & $0.049 $& $0.053 $  \\
$\text{II}_{\text{band}}$&$n_{\text{dop},2}$  & $0.047 $& $0.050 $  \\
&$n_{\text{dop},3}$ & $0.047 $& $0.049 $  \\
&&\\ \hline
&Undoped& $0.181 $& $0.126 $  \\
&$n_{\text{dop},1}$ & $0.142 $& $0.126 $  \\
$\text{III}_{\text{band}}$&$n_{\text{dop},2}$  & $0.140 $& $0.157 $  \\
&$n_{\text{dop},3}$ & $0.140 $& $0.199 $  \\
\end{tabular}
\caption{Effective masses for the hydrogenated (111) diamond surface in units of the electron mass $m_e$. Bands are labelled according to Fig.~\ref{fig:BO_D} (c).}
\label{tab:eff_mass_111}
\end{center}
\end{table}

\begin{figure}[h]
\centering
\subfloat[]
	{ \includegraphics[width=\linewidth]{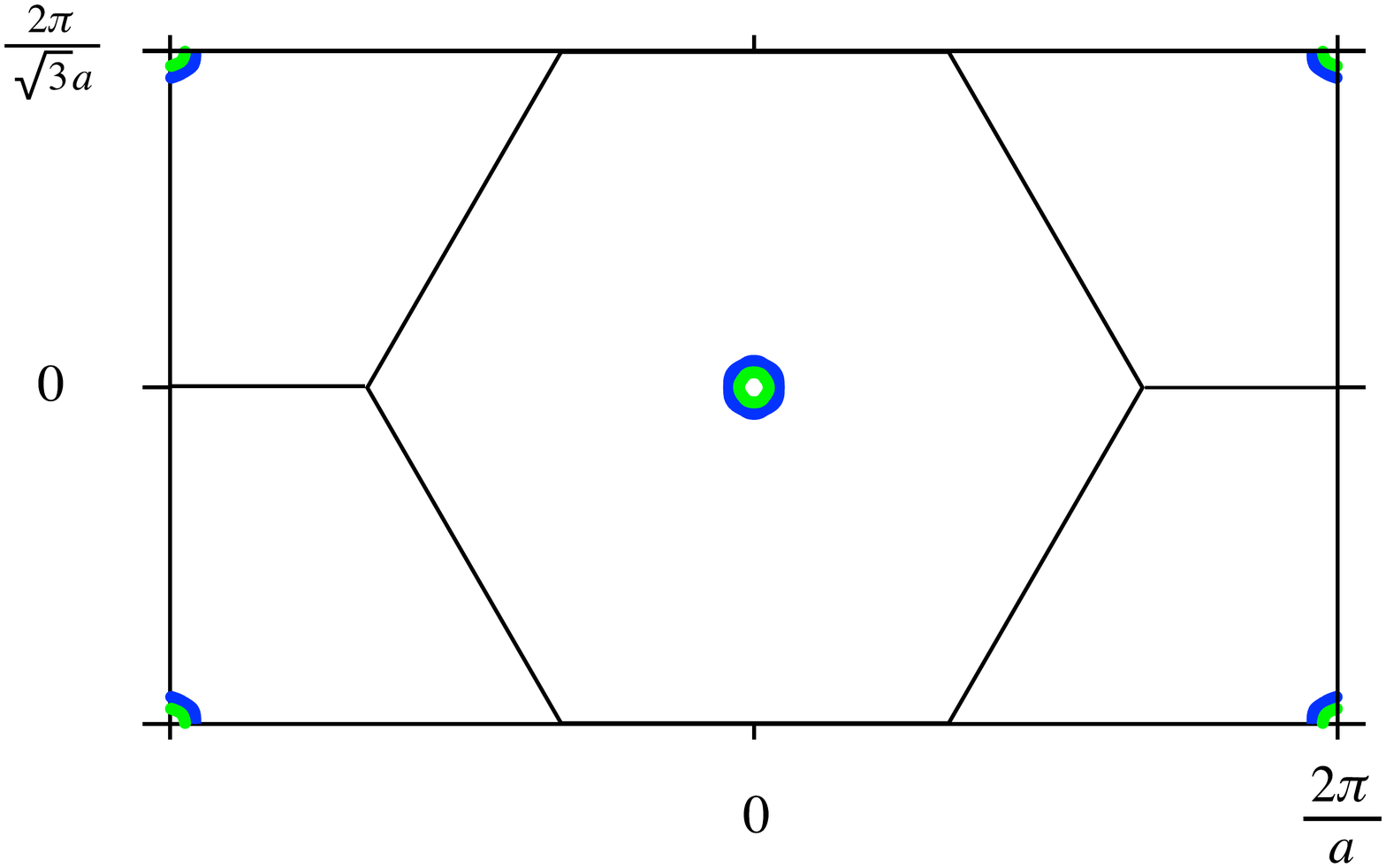}}\\
\subfloat[]
	{ \includegraphics[width=\linewidth]{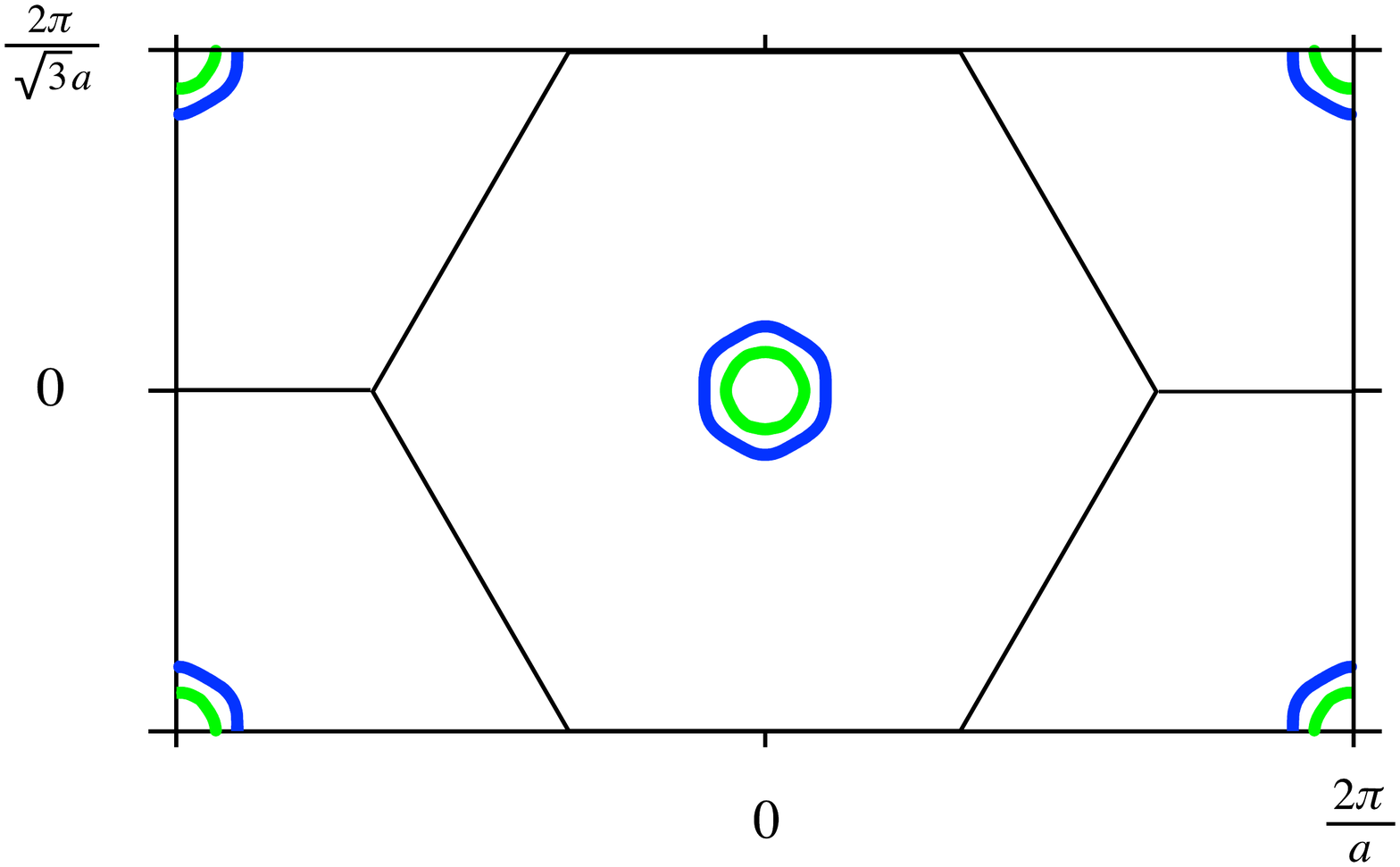}}\\
\subfloat[]
	{ \includegraphics[width=\linewidth]{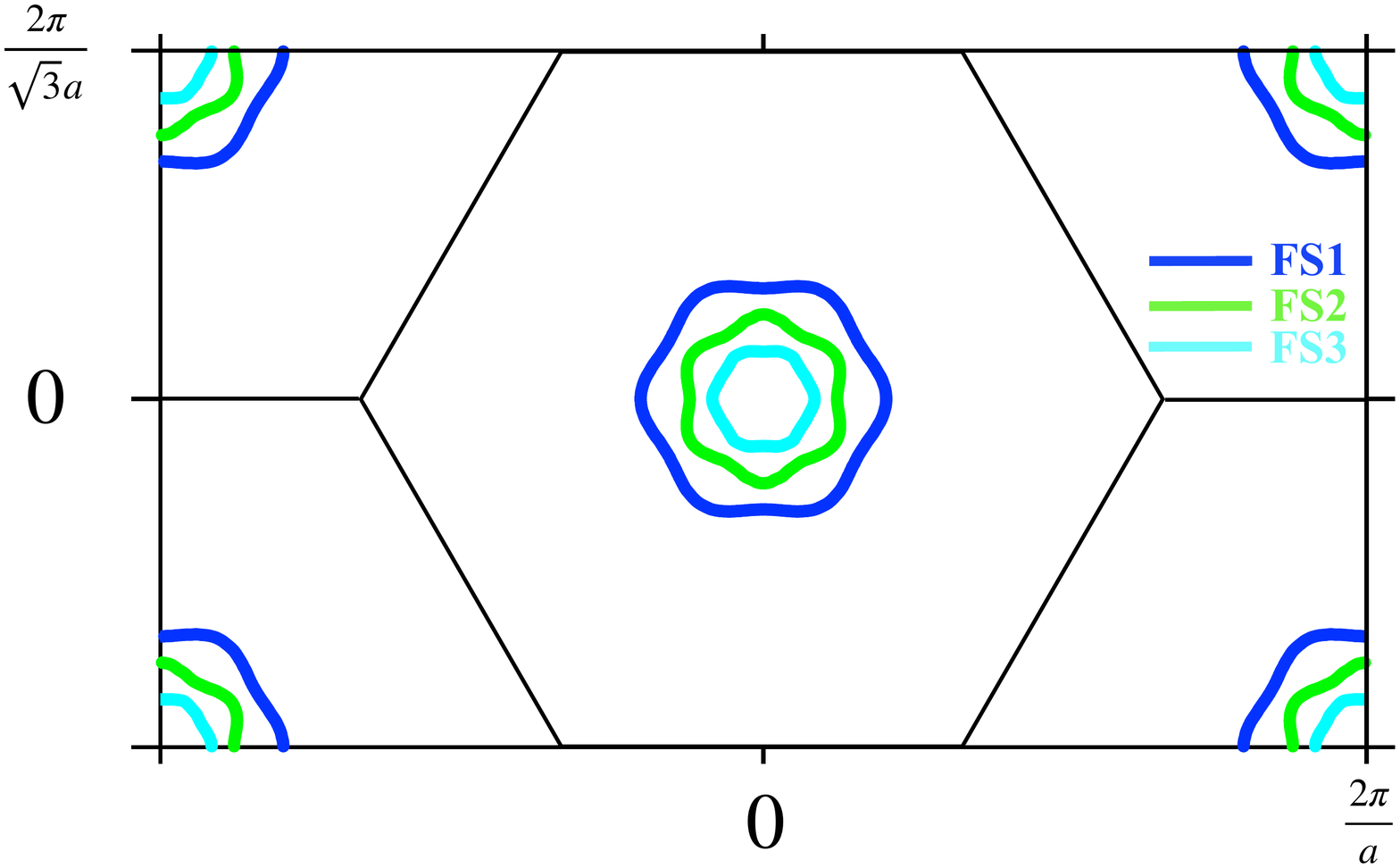}}\\	
 \caption
   {2D Fermi surfaces of the hydrogenated (111) diamond surface with (a) $n_{\text{dop},1}$, (b) $n_{\text{dop},2}$ and (c) $n_{\text{dop},3}$. FS1 $\rightarrow$ blue line, FS2 $\rightarrow$ green line and FS3 $\rightarrow$ aqua line.
   }
   \label{fig:111_Fermi}
\end{figure}

Let us finally discuss the evolution of the Fermi surface as a function of the induced charge density (Fig.~\ref{fig:111_Fermi}). The accumulation of carriers at the surface of our sample leads the system to a metallization of the first layers and, consequently, to the appearance of a Fermi surface whose extension varies as the chemical potential is lowered. Fermi surfaces labeling is depicted in Fig.~\ref{fig:111_Fermi} (c) where FS1 $\rightarrow$ blue line, FS2 $\rightarrow$ green line and FS3 $\rightarrow$ aqua line.

In the first case ($n_{\text{dop},1}$) two small concentric hole pockets appear at the center of the Brillouin zone (FS1 and FS2): the system has become a ``bad'' metal, in the sense that there are very few free carriers available for conduction. This corresponds to the situation observed by Yamaguchi et al.~\cite{Yamaguchi}, i.e. the sample is on the verge of an insulator-to-metal phase transition. By increasing the amount of holes at the surface, the Fermi surfaces remain centered at $\Gamma$ but grow in size, occupying a larger region of the Brillouin zone and signaling at least a metallization of the surface. Finally, at  $n_{\text{dop},3}$ the Fermi level crosses a third band generating a third hole pocket centered at $\Gamma$ (FS3). Another important aspect, clearly shown in Fig.~\ref{fig:111_Fermi}, is that the $C_{3v}$ hexagonal symmetry of the Fermi surfaces is not broken down by the presence of the strong electric field.

\subsection{\label{sec:vib}Vibrational and Superconductive Properties}

In this section we study how both the induced charge density and the presence of the electric field modify the vibrational properties of our system. First of all we analyze the phonon dispersion relations and density of states and then we move to the electron-phonon interactions at $\Gamma$ and, through Wannier interpolations, on the whole Brillouin zone at the maximum doping. Finally, in order to compute the superconductive transition temperature we use the Allen-Dynes/McMillan formula\cite{McMillan,Allen_Dynes}:

\begin{equation}
\label{eq:AD}
T_{\text{C}} = \frac{\omega_{\textrm{log}}}{1.2}\exp{-\frac{1.04(1+\lambda)}{\lambda - \mu^*(1+0.62\lambda)}}
\end{equation}
where $\lambda$ is the electron phonon coupling constant, $\omega_{log}$ is the logarithmic averaged frequency and $\mu^*$ is the Morel-Anderson pseudopotential\cite{Mor-And}. The first two quantities will be computed ab-initio through the following expressions:

\begin{equation}
\label{eq:lambda}
\lambda = 2\int d\omega \frac{\alpha^2F(\omega)}{\omega}
\end{equation}

\begin{equation}
\label{eq:omega_log}
\omega_{\textrm{log}} = \text{exp}\Bigl[\frac{2}{\lambda}\int\text{log}(\omega)\frac{\alpha^2F(\omega)}{\omega}d\omega\Bigr]
\end{equation}
while $\mu^*$ will be considered as a parameter ranging from 0.13 to 0.14 (which are the values used in boron-doped diamond~\cite{Blase}). The quantity $\alpha^2F(\omega)$ is the Eliashberg spectral function and represents the spectral decomposition of the electron-phonon coupling constant $\lambda$:

\begin{equation}
\label{eq:Eliashberg_function}
\begin{split}
\alpha^2F(\omega) = & \frac{1}{N_{\text{tot}}(0)N_qN_k}\sum_{\vb{q}\nu}\delta(\hbar\omega-\hbar\omega_{\vb{q}\nu})\cdot\\ & \cdot\sum_{\vb{k},n,m}\abs{g_{\vb{k}n,\vb{k}+\vb{q}m}^{\nu}}^2\delta(\epsilon_{\vb{k}n})\delta(\epsilon_{\vb{k}+\vb{q}m})
\end{split}
\end{equation}
where $N_{\text{tot}}(0)$ is the total density of states per spin, $N_q$ and $N_k$ are the number of q- and k-points in the first Brillouin zone considered for the computations, and the energy $\epsilon$ is measured from the Fermi level. The electron-phonon matrix elements $g_{\vb{k}n,\vb{k}+\vb{q}m}^{\nu}$ between band $n$ and $m$ ($n,m=1,3$ in our case, as labelled in Fig.~\ref{fig:111_Fermi}) for phonon mode $\nu$ are defined as:

\begin{equation}
g_{\vb{k}n,\vb{k}+\vb{q}m}^{\nu} =\sum_{A\alpha} \frac{e^{A\alpha}_{\vb{q}\nu}}{\sqrt{2M_A\omega_{\vb{q}\nu}}}\bra{\vb{k}n}\fdv{v_{\text{SCF}}}{u_{A\alpha}^{\vb{q}}}\ket{\vb{k}+\vb{q}m} \label{eq:matrixelements}
\end{equation}
where $\ket{\vb{k}n}$ is the Bloch-periodic part of the Kohn-Sham eigenfuction, $v_{\text{SCF}}=e^{-i\vb{q}\cdot\vb{r}}V_{\text{KS}}$ is the periodic part of the Kohn-Sham potential $V_{\text{KS}}$. Atoms in the unit cell are labelled by $A$ and their cartesian coordinates by $\alpha$, while their mass is denoted by $M_A$. The Fourier transformed displacement of atom A along the cartesian direction $\alpha$ is denoted by $u_{A\alpha}^{\vb{q}}$ and $e^{A\alpha}_{\vb{q}\nu}$ is the phonon eigenvector normalized on the unit cell of components $A\alpha$.\\

As we discussed above, on increasing the doping the system evolves towards a quasi-2D structure, where the use of Eq.~\ref{eq:AD} is not completely justified. In fact, in the 2D limit, fluctuations are greatly renormalized\cite{fluctuations} and the long-range order of electron-phonon interaction might be destroyed leading to other coupling mechanisms\cite{BKT}. However, we will still use  eq. \ref{eq:AD} to estimate $T_{\text{C}}$, keeping in mind that the result should be considered as an upper bound for the actual superconducting critical temperature. \\

\subsubsection{\label{sec:Vib}Phonon dispersion relations and DOS}

\begin{figure}
\centering
\subfloat[]
	{ \includegraphics[width=\linewidth]{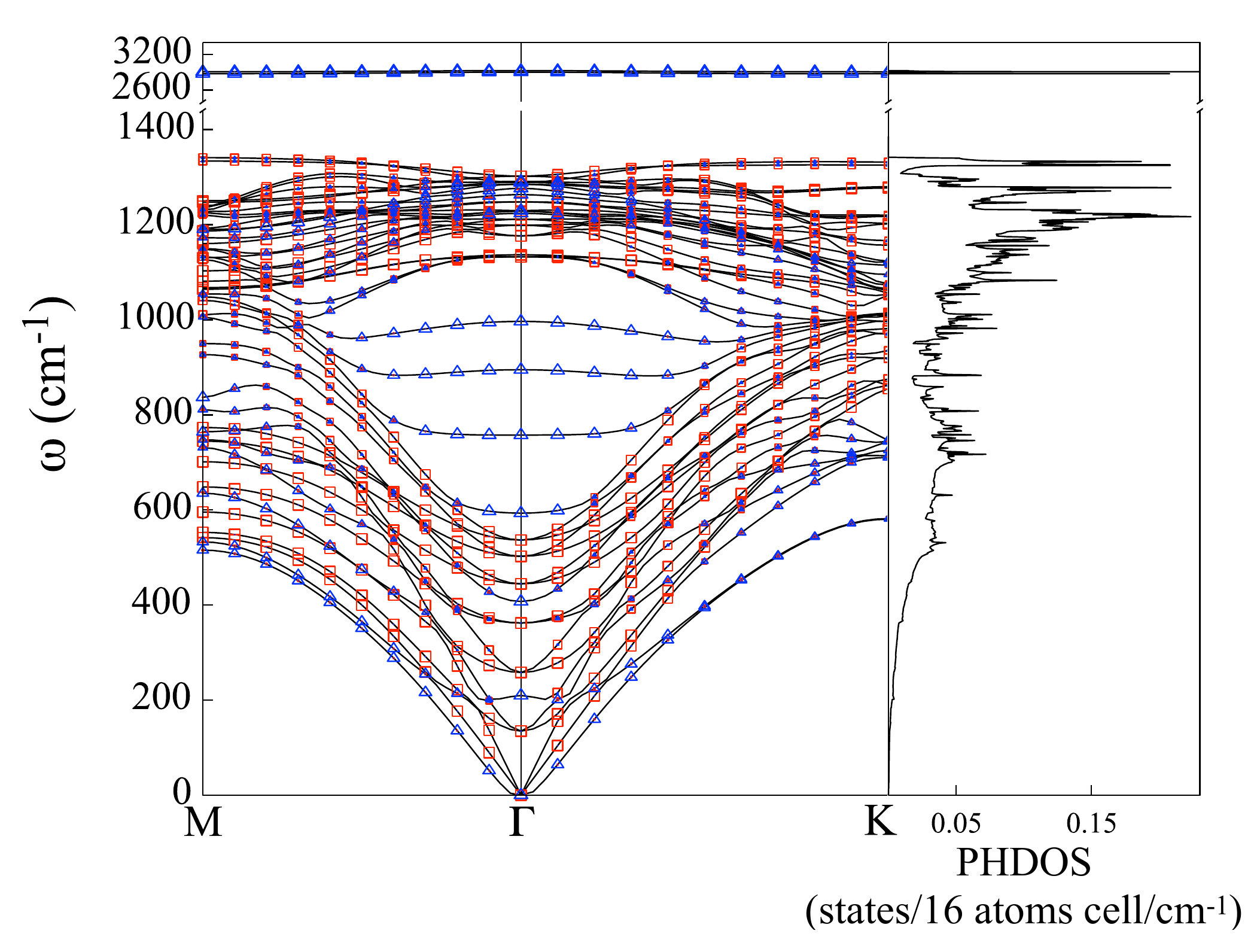}}\\
\subfloat[]
	{ \includegraphics[width=\linewidth]{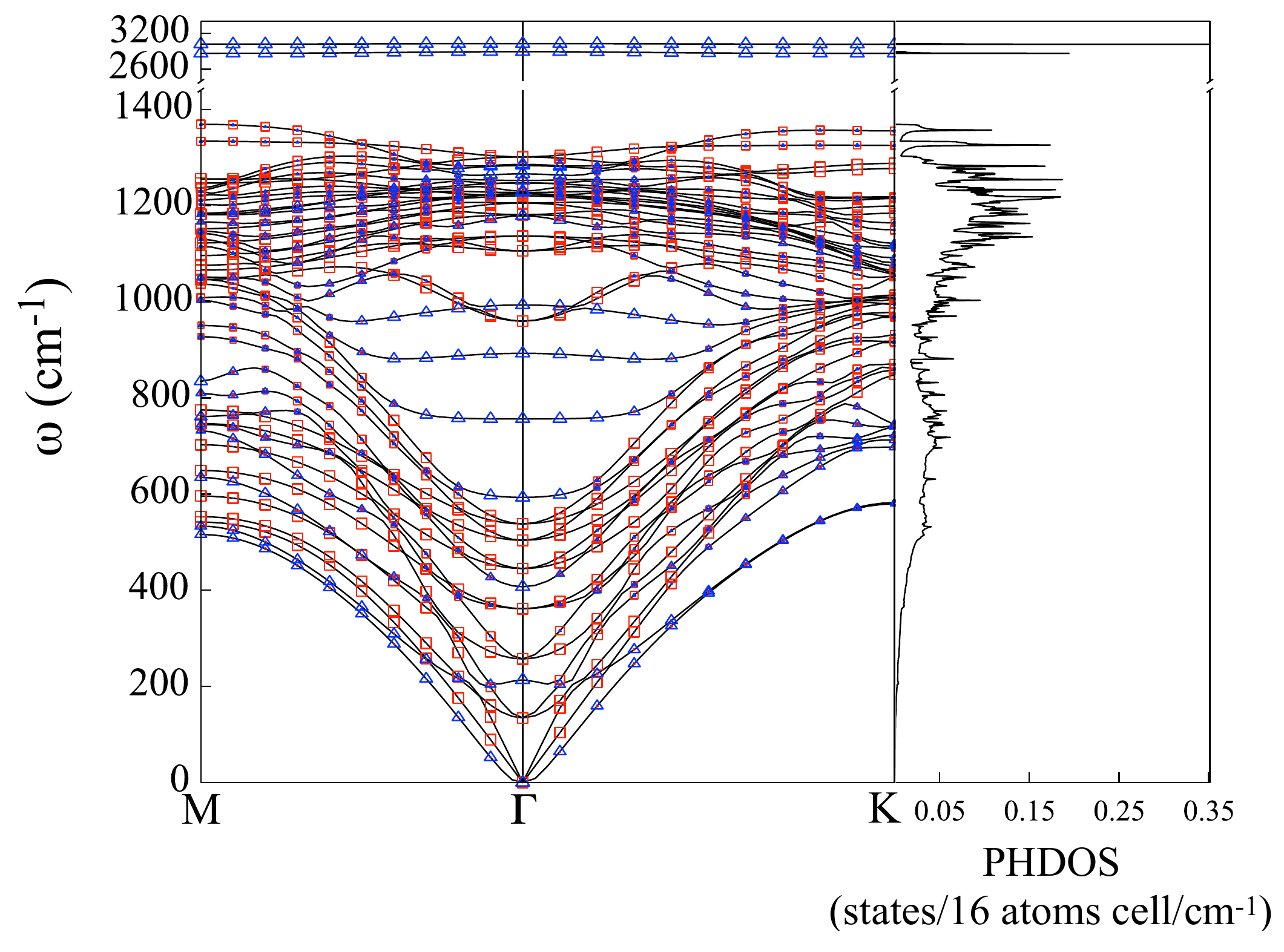}}\\
\subfloat[]
	{ \includegraphics[width=\linewidth]{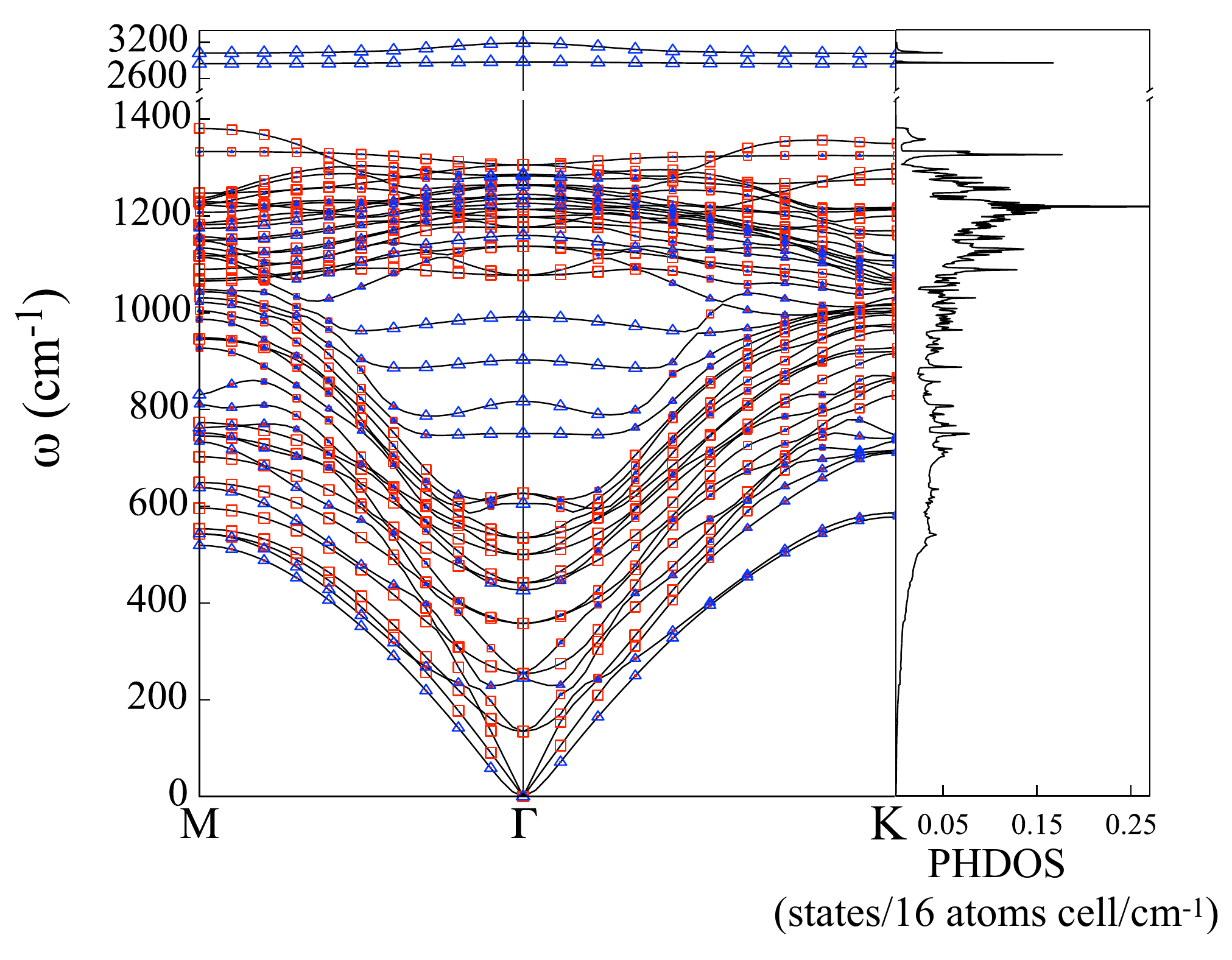}}\\	
 \caption
   {Phonon dispersion relations and density of states (DOS) of the hydrogenated (111) diamond surface with (a) $n_{\text{dop},1}$, (b) $n_{\text{dop},2}$ and (c) $n_{\text{dop},3}$. Red (blue) dots correspond to in-plane (out-of-plane) vibrational modes. The size of the dots correspond to the amount of [xy] or [z] nature of the phonon frequency. For example, the size of the dots for the modes at $\Gamma$ corresponds to a $100\%$ character, either [xy] or [z].}
   \label{fig:111_PD_G}
\end{figure}

In Fig.~\ref{fig:111_PD_G} we plot the vibrational dispersion relations of our system and the phonon density of states for the three doping values considered in the paper. The in-plane and out-of plane nature of the displacement eigenvalues is indicated by red and blue dots respectively, whose size varies according to the percentage of $[xy]$ and $[z]$ character at a specific $\vb{q}$ point in the Brillouin zone for each mode $\nu$. At  $\sim3000$ cm$^{-1}$ there are the two out-of-plane optical branches of the hydrogen atoms which are degenerate at low doping. On increasing the applied electric field, the degeneracy is lifted giving rise to two distinct high energy flat modes with very peaked density of states. The in-plane hydrogen modes, instead, are already greatly softened at the first doping $n_{\text{dop},1}$ (see Fig.~\ref{fig:111_PD_G}a): indeed in this case there are two degenerate modes at $\sim1132.40$ cm$^{-1}$ and other two degenerate modes at $1136.69$ cm$^{-1}$.

Carbon atoms frequencies instead range from $0$ to $\sim1400$ cm$^{-1}$ and their eigenvalues arrange in very rich dispersion relations. For all three doping values, the phonon DOS for the carbon atoms is very concentrated around high-energy modes (where we have a majority of nearly-flat bands) while the low energy ones are less populated.

\begin{figure}
\centering
\subfloat[]
	{\hspace*{-1.5cm}
	\includegraphics[width=.87\linewidth]{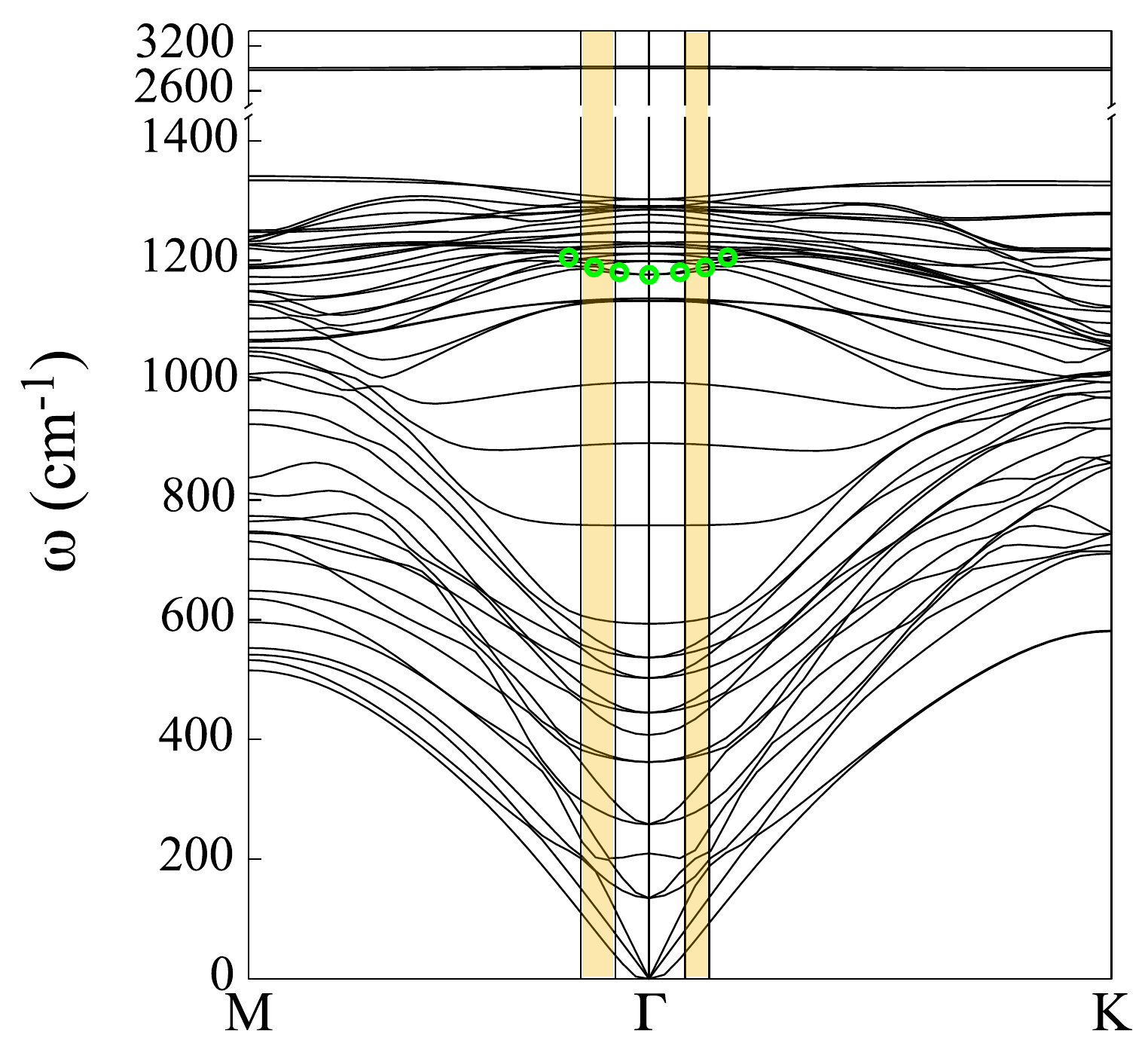}}\\
\subfloat[]
	{\hspace*{-1.5cm}
	 \includegraphics[width=.9\linewidth]{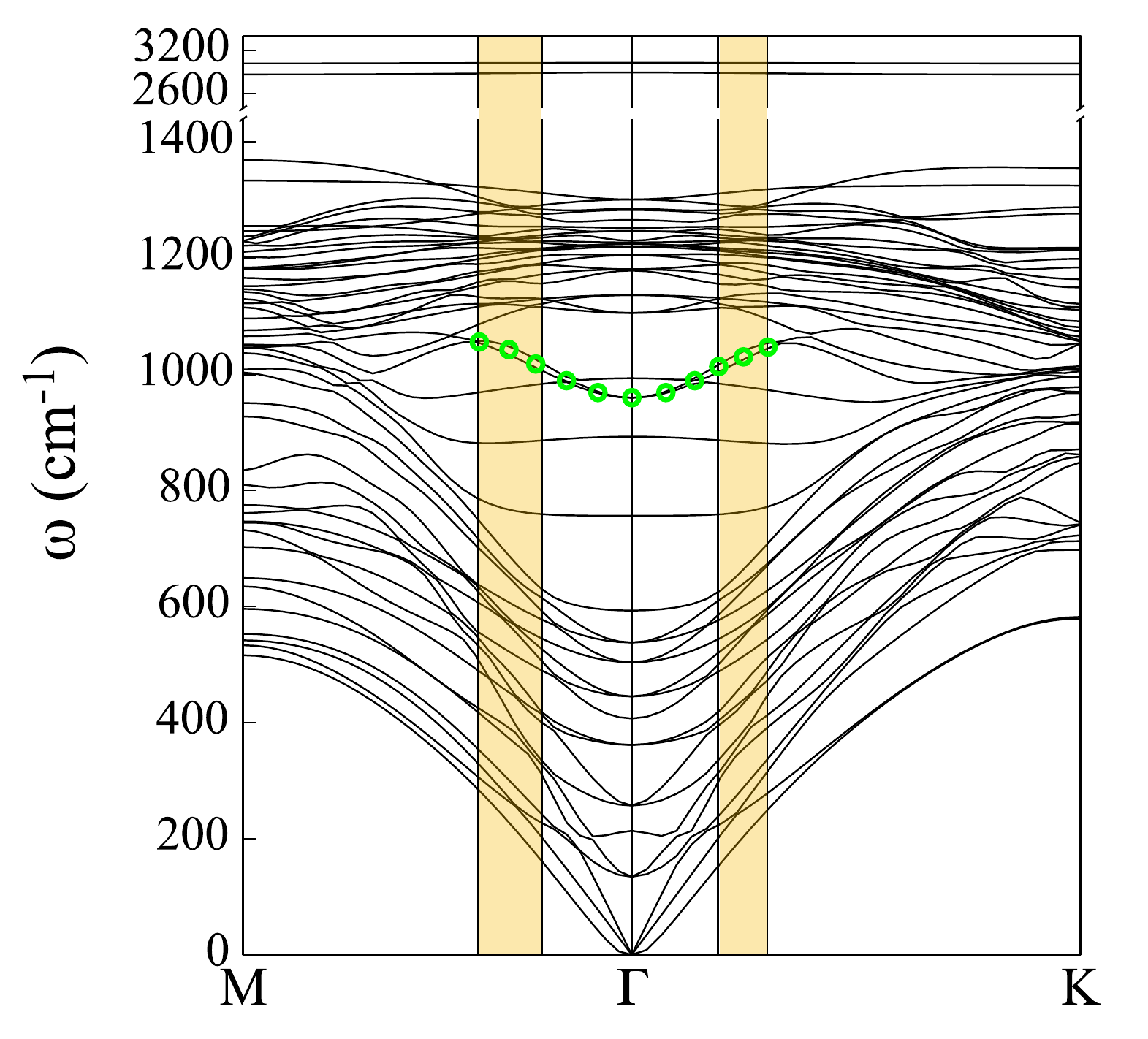}}\\
\subfloat[]
	{\hspace*{-1.5cm}
	\includegraphics[width=.85\linewidth]{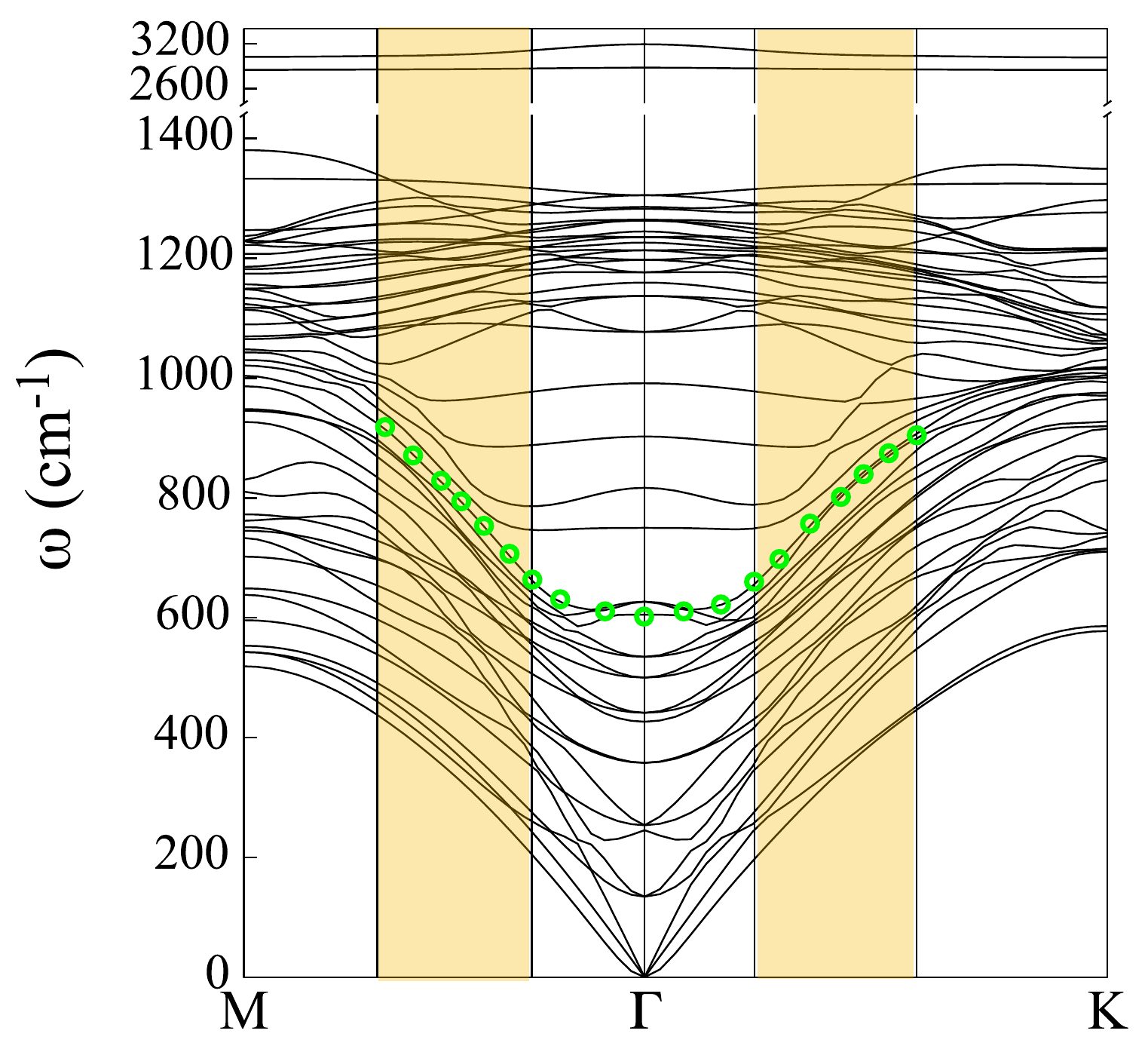}}\\	
 \caption
   {Softening of phonon modes of the hydrogenated (111) diamond surface with (a) $n_{\text{dop},1}$, (b) $n_{\text{dop},2}$ and (c) $n_{\text{dop},3}$. Green dots highlight the band which is actually being renormalized. Orange regions comprise the momenta interval $\vb{q}\in[2\vb{k}_{F,\text{min}},2\vb{k}_{F,\text{max}}]$, where $2\vb{k}_{F,\text{min}}$ correspond to the smallest Fermi surface while $2\vb{k}_{F,\text{max}}$ correspond to the biggest one.}
   \label{fig:111_soft_G}
\end{figure}

From the phonon dispersion relation it is also possible to observe the softening of an optical in-plane mode, which is highlighted in Fig.~\ref{fig:111_soft_G} by green circles. As one goes from low to high doping values, the frequencies of this branch become lower and lower: the system develops a Kohn anomaly\cite{Kohn_an}, that is a renormalization of the phonon modes due to the electron-phonon interaction, which occurs in a region of radius $2k_F$ around $\vb{\Gamma}$ and has been observed in boron-doped diamond\cite{Boeri, Giustino1} and p-doped graphane\cite{Giustino2}. In order to see this effect we make the approximation of circular Fermi surfaces of radius $\vb{k}_F$; in our case there are up to three bands crossing the Fermi level and, consequently, up to three Fermi momenta. In Fig.~\ref{fig:111_soft_G} the orange bands highlight the region such that $|q|\in[2\|\vb{k}_{F,\text{min}}\|,2\|\vb{k}_{F,\text{max}}\|]$, where $\vb{k}_{F,\text{min}}$ and $\vb{k}_{F,\text{max}}$ are the wavevectors corresponding to the smallest and largest Fermi surface, respectively. In principle, a change in slope of the phonon branches due to the Kohn anomaly should be observed in correspondence of \emph{each} Fermi momentum, but this would require an extremely fine discretization of the reciprocal space. However, even with our choice of k-points, a slope change of the aforementioned mode can be observed within the orange regions. It is very small for low doping values, but becomes evident as the maximum doping regime is reached (see Fig.~\ref{fig:111_soft_G}(c)). This clearly indicates a strong effect of the electron-phonon interactions on the properties of this system.

\subsubsection{\label{sec:Gamma}Electron-Phonon Computation at $\vb{q}=\vb{\Gamma}$}

\begin{figure*}
\centering
\subfloat[]
	{ \includegraphics[width=0.4\linewidth]{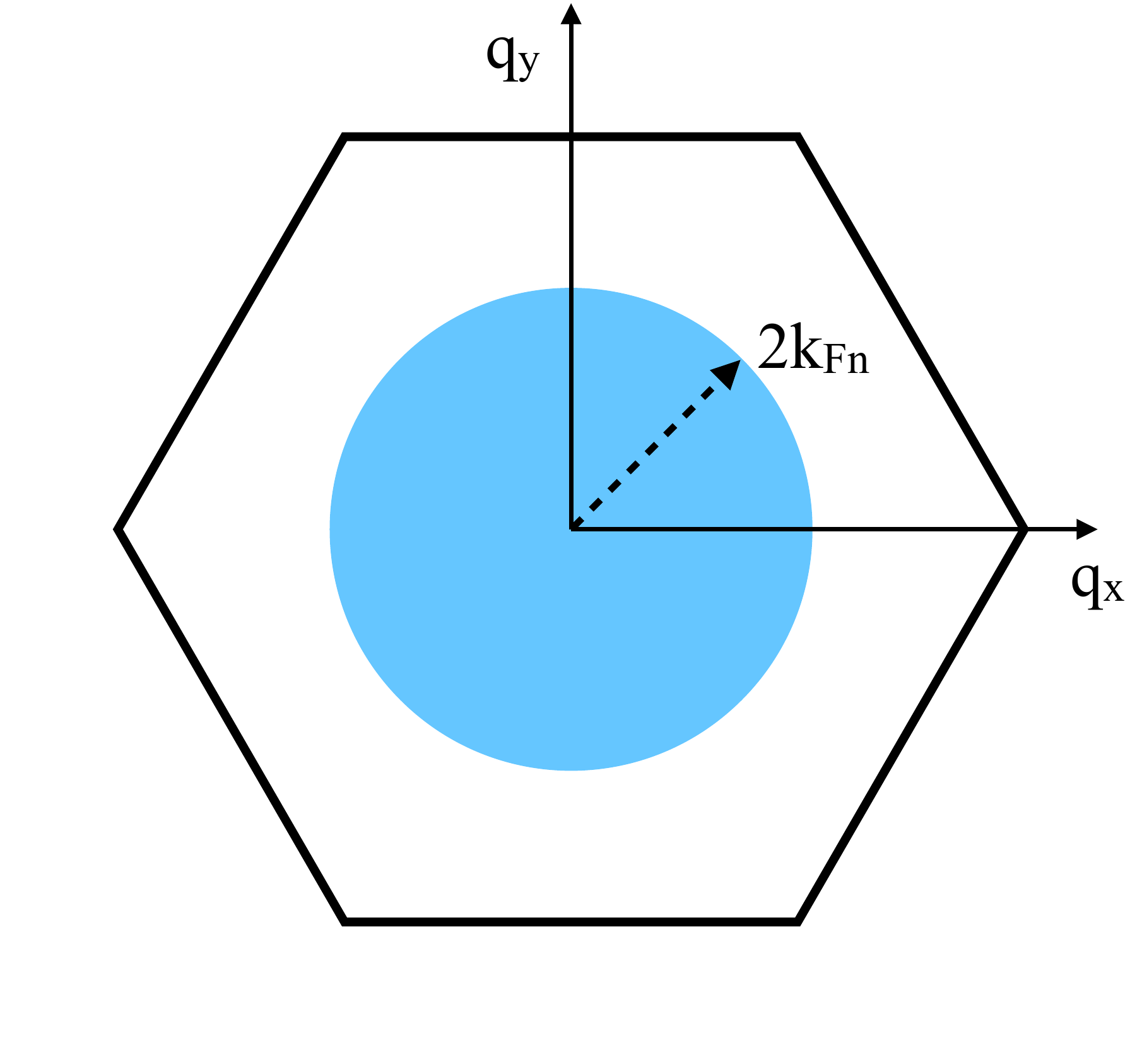}}
\subfloat[]
	{ \includegraphics[width=0.4\linewidth]{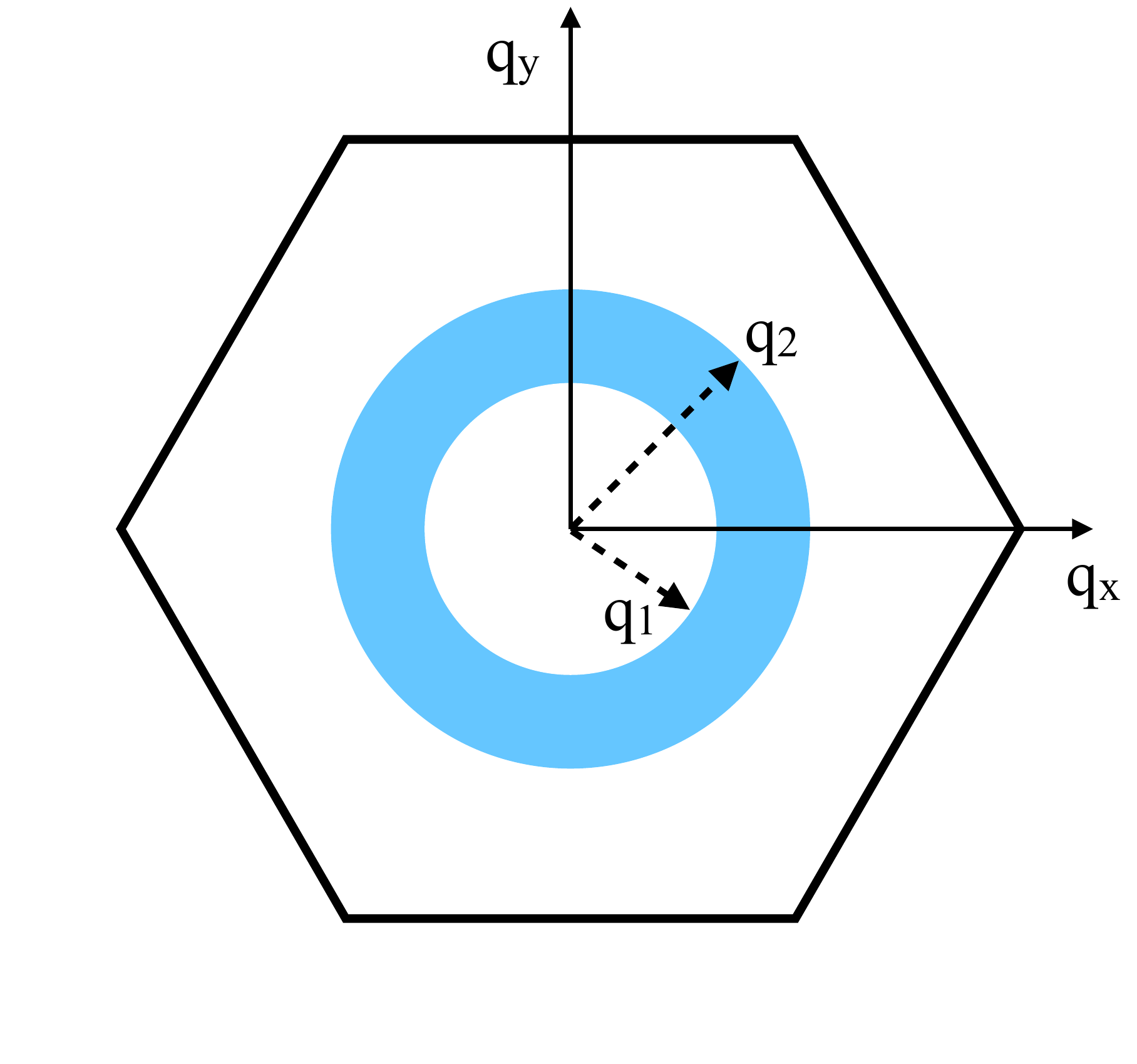}}\\
 \caption
   {$\vb{q}$-space portions of the Brillouin zone (blue regions) over which the electron-phonon matrix elements are considered to be constant (and thus can be represented by their value at at $\vb{q}=0$) in the case of  $n=m$ (a) and $n\ne m$ (b).}
   \label{fig:model_gamma}
\end{figure*}

Let us start with the calculation of electron-phonon interactions at $\vb{q}=\Gamma$, which is a computationally inexpensive task. This procedure provides approximate results with the assumption that the electron-phonon matrix elements (eq. \ref{eq:matrixelements}) are constant, or nearly constant, in the portion of reciprocal space delimited by the phonon vectors $\vb q$ that give non-zero contributions to the nesting factor:

\begin{equation}
\label{eq:nesting}
N_f(\vb{q})=\frac{1}{N_k}\sum_{\vb{k},nm}\delta(\epsilon_{\vb{k}n})\delta(\epsilon_{\vb{k}+\vb{q}m})
\end{equation}

The Dirac deltas appearing in eq.~\ref{eq:nesting} limit the summation over the vectors $\vb{q}$ which can scatter an electron from the $n$-th to the $m$-th Fermi surface (since we have set $E_F=0$ in Eq.~\ref{eq:nesting}). As a matter of fact, if we define $\vb{k}'=\vb{k}+\vb{q}$, the first Delta tells us that $\abs{\vb{k}}=k_{Fn}$ while the second one gives $\abs{\vb{k}'}=k_{Fm}$. Therefore, since $\vb{q}=\vb{k}'-\vb{k}$ we have that:

\begin{equation}
\begin{split}
\abs{\vb{q}}=&\sqrt{\abs{\vb{k}}^2+\abs{\vb{k}'}^2-2\abs{\vb{k}}\abs{\vb{k}'}\cos{\theta}}=\\
=&\sqrt{k_{Fn}^2+k_{Fm}^2-2k_{Fn}k_{Fm}\cos{\theta}}
\end{split}
\end{equation}
where $\theta$ is the angle between $\vb{k}$ and $\vb{k}'$. \\
Then, assuming spherical Fermi surfaces, the electron-phonon matrix elements computed at $\vb{q}=\vb{\Gamma}$ are considered constant inside the region delimited either by a circle of radius $\abs{\vb{q}}=2k_{Fn}$ if $n=m$ (Fig.~\ref{fig:model_gamma}(a)) or by an annulus of radii $\abs{\vb{q}_1}=\sqrt{k_{Fn}^2+k_{Fm}^2-2k_{Fn}k_{Fm}}$ and $\abs{\vb{q}_2}=\sqrt{k_{Fn}^2+k_{Fm}^2+2k_{Fn}k_{Fm}}$ if $n\ne m$ (Fig.~\ref{fig:model_gamma}(b)).\\

In order to see which are the strongest electron-phonon modes as a function of the induced charge density, we compute the squared average of the electron-phonon matrix elements for each phonon mode $\nu$:

\begin{equation}
\langle g_{\nu}^2 \rangle_{\Gamma} =\sum_{n,m}\frac{\abs{g_{\Gamma n,\Gamma m}^\nu}^2N_n(0)N_m(0)}{N_{tot}^2(0)}
\end{equation}
where $N_{n}(0)$ is the density of states of band $n$ at the Fermi level, $n,m=1,2,3$ are band indices (obtained from Fig.~\ref{fig:111_Fermi}) and $\nu$ is the mode index.

In the case of $n_{\text{dop},1}$, there are two degenerate in-plane modes at $\omega_{\shortparallel} = 1176$ cm$^{-1}$ that have the strongest e-ph matrix elements, with $\langle g_{\shortparallel}^2 \rangle = 0.167$ eV$^2$. These modes mainly contribute to intraband $1-1$ and interband $1-2$ scattering processes, while the $2-2$ channel is negligible.

As for $n_{\text{dop},2}$ there are again two degenerate in-plane modes at $\omega_{\shortparallel} = 959.25$ cm$^{-1}$ with $\langle g_{\shortparallel}^2 \rangle = 0.441$ eV$^2$. In this case the in-plane mode contributes both to inter- and intra-band processes which are relevant for the electron-phonon interaction, including the $2-2$ scattering process. This is reasonable since at this doping level the size of the second band is no longer negligible.

At the highest doping level, i.e. $n_{\text{dop},3}$, the largest $\langle g_{\nu}^2 \rangle$ are associated to two degenerate in-plane modes at $\omega_{\shortparallel} = 626$ cm$^{-1}$ (with $\langle g_{\shortparallel}^2 \rangle = 1.105$ eV$^2$) and a single out-of-plane mode at at $\omega_{\bot} = 817$ cm$^{-1}$ (with $\langle g_{\bot}^2 \rangle = 0.248$ eV$^2$). In this case the only negligible process of the in-plane mode is the intra-band scattering within the third Fermi surface. On the other hand, the out-of-plane mode only favors the intra-band processes of all the three bands. \\
The fact that also an out-of-plane mode becomes relevant in this last case is not so obvious. However, the electronic projected density of states of Fig.~\ref{fig:111_pdos} clearly shows that the first two bands have only a planar nature, and the out-of-plane character only appears when the third band crosses the Fermi level. Moreover, the strongest in-plane mode is actually the one which is being softened by the Kohn anomaly (Fig.~\ref{fig:111_soft_G}) discussed above: as the doping value increases, the average in-plane electron-phonon interaction increases and this translates in a strong renormalization of the phonon frequencies.\\

\begin{figure}
\centering
   \includegraphics[width=1.05\linewidth]{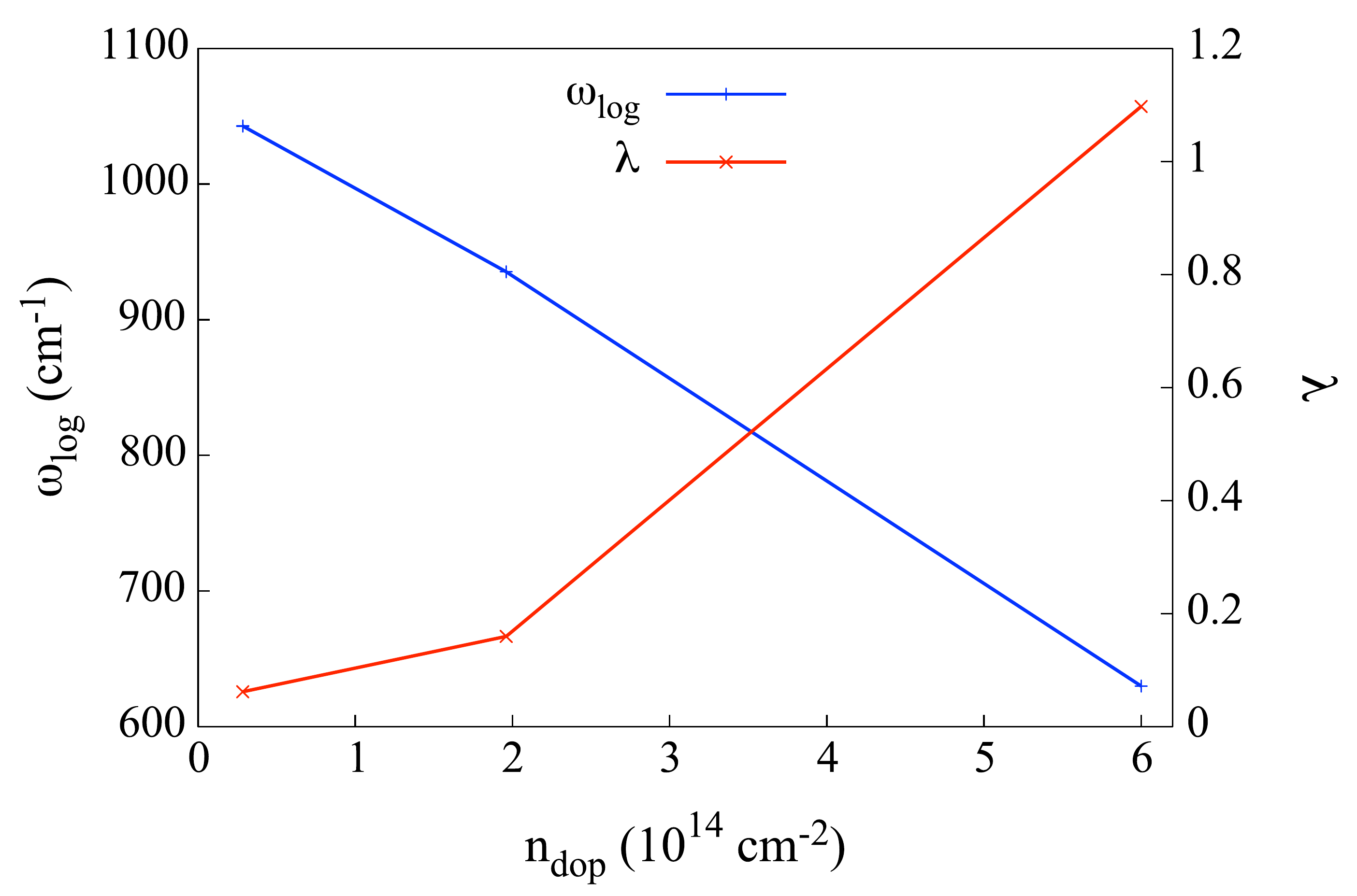}
   \caption
 {Superconductive parameters ($\omega_{log}$ and $\lambda$) for the hydrogenated (111) diamond surface as a function of the induced charge density ($n_{\text{dop}}$) computed in the $\vb{q}=\Gamma$ approximation.
   }
   \label{fig:111_SG}
\end{figure}

With these pieces of information we can compute $\lambda$ and $\omega_{\textrm{log}}$ (Fig.~\ref{fig:111_SG}) and then estimate the superconducting critical temperature $T_{\text{C}}$. The electron-phonon coupling constant $\lambda$ is just an average of the electron-phonon matrix elements $g_{\Gamma n,\Gamma m}^\nu$ over the Brillouin zone weighted by the vibration eigenvalue $\omega_{\Gamma\nu}$, i.e. the frequency of the phonon of mode $\nu$ at $\vb{q}=\vb{\Gamma}$:

\begin{equation}
\label{eq:lambda_G}
\begin{split}
\lambda(\Gamma) = &\sum_{n,m}\lambda_{nm}(\Gamma)\\
= &\sum_{n,m} \Biggl\{\sum_{\nu}\frac{\abs{g_{\Gamma n,\Gamma m}^\nu}^2}{\omega_{\Gamma\nu}}\frac{N_n(0)N_m(0)}{N_{\text{tot}}(0)}\Biggr\}
\end{split}
\end{equation}

The logarithmic averaged frequency $\omega_{\textrm{log}}$ is given instead by:

\begin{equation}
\label{eq:w_G}
\omega_{\textrm{log}}(\Gamma) = \exp\Biggl[\frac{\sum_{\nu,n,m}\frac{\log{\omega_{\Gamma\nu}}\abs{g_{\Gamma n,\Gamma m}^\nu}^2}{\omega_{\Gamma \nu}}N_n(0)N_m(0)}{\sum_{\nu,n,m}\frac{\abs{g_{\Gamma n,\Gamma m}^\nu}^2 }{\omega_{\Gamma\nu}}N_n(0)N_m(0)}\Biggr]
\end{equation}


In Fig.~\ref{fig:111_SG} we plot the evolution of these two quantities as a function of doping.
As we increase the amount of the induced charge, the logarithmic averaged phonon frequency decreases while the electron-phonon coupling constant increases.The increase in $\lambda$ can be understood as being due to the softening of the phonon modes with the strongest matrix elements. As a matter of fact, in the calculation of $\lambda$ (see eq.\ref{eq:lambda}) the e-ph interactions occurring at low frequencies are enhanced with respect to those at high frequency. When the phonon modes that give peaks in $\alpha^2 F(\omega)$ move to lower frequencies, their contribution to the coupling constant is amplified and $\lambda$ increases from $\approx0.06$ (at $n_{\text{dop},1}$) to $\approx1.09$ (at $n_{\text{dop},3}$). On the other hand, the averaged logarithmic frequency ($\omega_{\textrm{log}}$) decreases upon doping, since phonon modes are being softened. Therefore there is a trade-off between $\lambda$ and $\omega_{\mathrm{log}}$.\\

The calculated values of $\lambda$ and $\omega_{\mathrm{log}}$ can now be inserted in the Allen-Dynes formula to determine $T_{\text{C}}$. It turns out that, for the first two doping values ($n_{\text{dop},1}$ and $n_{\text{dop},2}$) there is no superconducting transition. However at $n_{\text{dop},3}$ there is a superconductive phase with a critical temperature in the range $63.14-57.20$ K for $\mu^*\in[0.13;0.14]$. The corresponding value of $\lambda$ is almost three times the value of boron-doped diamond\cite{Blase} ($\lambda_{B}=0.43$) and this is reasonable since the local density of holes on the first two carbon layers exceeds by an order of magnitude the boron concentration responsible for a $T_{\text{C}}=4$ K. The increase in the electron-phonon interaction, with respect to the previous two doping values, can be related to the fact that, when the third band crosses the Fermi level, the density of states doubles. However, this may also be an artifact of the simple model at $\vb{q}=\Gamma$ and therefore more precise computations are required for this last case.\\

\subsubsection{\label{sec:Wann}Electron-Phonon Computations through Wannier interpolation}

Since the highest doping value ($n_{\text{dop},3}$) seems to show a superconductive phase transition with a sizable critical temperature, we try to further investigate this phenomenon by a more accurate procedure. This is accomplished by interpolating the electron-phonon matrix elements over the whole Brillouin zone by using the Wannier functions. In this way we do not consider the interactions only at $\vb{q}=\vb{\Gamma}$, but we actually compute their value for every point in reciprocal space.

Since the atoms which contribute most to the DOS at the Fermi level are the carbon atoms belonging to the first three layers of the samples, in order to fit the bands which cross the Fermi energy we use 3 sp$^3$ orbitals centered on C(1), C(2) and C(3) for a total of 12 Wannier functions. The resulting Wannier bands are quite satisfactory, since the Fermi velocities (i.e. the derivative of the energy bands computed at the Fermi level) of the Wannierized bands are in good agreement with that of the original bands. \\

\begin{table}
\begin{center}
\begin{tabular}{c|c|c|c}
$ $& $\lambda$ & $\omega_{log}$ (cm$^{-1}$) & $T_{\text{C}}$ (K) \\
\hline
$\Gamma$ &$1.09$ & $629.94$ & $63.14-57.20$ \\
Wannier & $0.81$ & $670.17$ & $34.93-29.60$ \\
\hline
\end{tabular}
\caption{Comparison between the electron-phonon coupling constant $\lambda$ and logarithmic averaged phonon frequency computed only at $\Gamma$ and obtained by Wannier interpolation on the whole Brillouin zone for the case at the highest doping..}
\label{tab:wann_vs_G}
\end{center}
\end{table}

The calculation of the electron-phonon interaction performed by using the Wannier functions gives an electron-phonon coupling constant $\lambda=0.81$, which is $\sim 35\%$ smaller than that computed at $\vb{q}=\vb{\Gamma}$. However, the logarithmic averaged frequency obtained through Wannierization is bigger than that obtained via the simplified approach, with a difference of $\sim 40$ cm$^{-1}$. This reduces the final critical temperature to $T_{\text{C}}\in[34.93-29.60]$ K for $\mu^*\in[0.13;0.14]$, which is $\sim30$ K smaller than the one obtained by diagonalization of the dynamical matrices computed only at the center of the Brillouin zone. This means that, from a quantitative point of view, the simple model we adopted before overestimates the critical temperature, but qualitatively it seems to catch the main physics of this system. A comparison of the quantities relevant for the superconductivity is reported in Tab.~\ref{tab:wann_vs_G}.

\begin{figure}
\centering
\includegraphics[width=\linewidth]{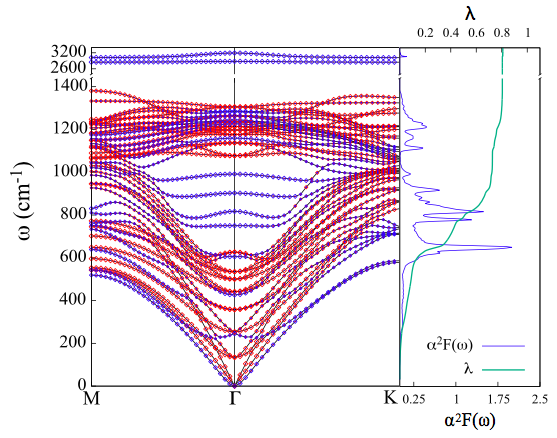}
 \caption{Phonon dispersion relations, Eliashberg spectral function $\alpha^2F(\omega)$ and the electron-phonon coupling constant $\lambda$ of the hydrogenated (111) diamond surface with $n_{\text{dop},3}$. Red (blue) dots correspond to in-plane (out-of-plane) vibrational modes. The size of the dots correspond to the amount of [xy] or [z] nature of the phonon frequency.}
   \label{fig:111_wann_disp}
\end{figure}

The Wannier interpolation can actually give us further details on the nature of the phase transition. Fig.~\ref{fig:111_wann_disp} reports the in-plane and out-of-plane components of the phonon dispersion relation (left panel), the Eliashberg spectral function $\alpha^2F(\omega)$ and the electron-phonon coupling constant $\lambda$ (right panel) resulting from the Wannierization of the bands. The $\alpha^2F(\omega)$ has two main peaks in correspondence of an in-plane phonon mode, which is the one that has been softened due to the Kohn anomaly, and of an out-of-plane vibrational mode. These peaks are responsible for the increase of the electron-phonon coupling constant since $\lambda$ has two clear jumps in correspondence of these modes. This is in agreement with what we have seen before: these are the phonon modes which the strongest interactions with electrons. However, from a quantitative point of view, the Wannier analysis shows that it is the out-of-plane mode the one with the highest matrix elements, contrary to what found with the simple $\vb{q}=\vb{\Gamma}$ calculation.

\begin{figure}
\centering
   \includegraphics[width=1.0\linewidth]{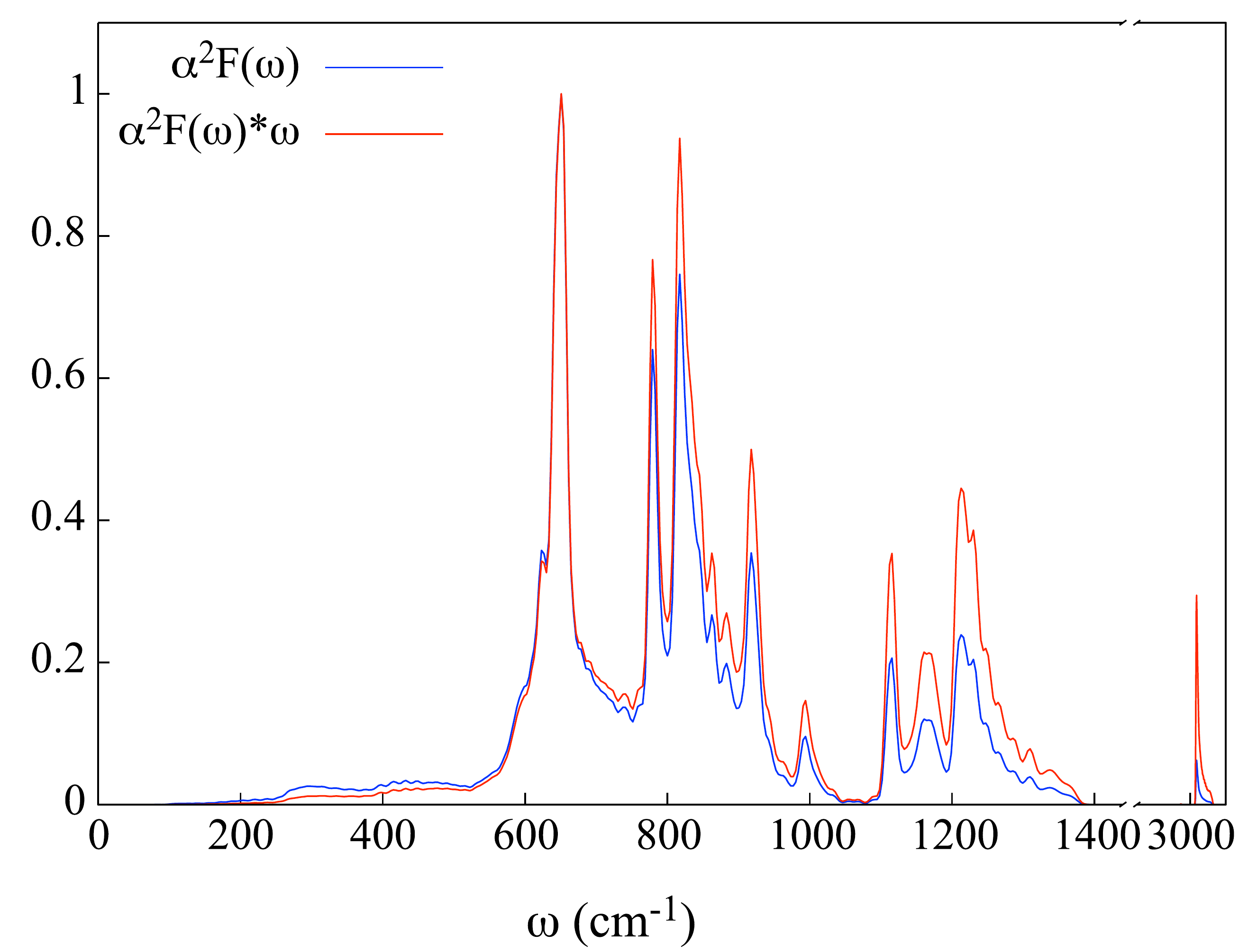}
   \caption
 {Comparison between the normalized Eliashberg spectral function $\alpha^2F(\omega)$ and its normalized first momentum $\alpha^2F(\omega)\cdot\omega$ for the case at the highest doping.
   }
   \label{fig:111_a2F_cnf}
\end{figure}

Since the Eliashberg spectral function is inversely proportional to $\omega$, we can plot $\alpha^2F(\omega)\cdot\omega$ in order to get rid of its the frequency dependece. In Fig.~\ref{fig:111_a2F_cnf} we plot both the normalized Eliashberg spectral function and its normalized first momentum. We can see that the peak occurring at the lowest frequency is actually unperturbed by the $1/\omega$ behavior, while the second peak is softened by this frequency dependence. Therefore their role in enhancing the electron-phonon interaction is comparable and we remind that this happens when we cross the third band.

\begin{figure}
\centering
\hspace*{-0.75cm}
   \includegraphics[width=\linewidth]{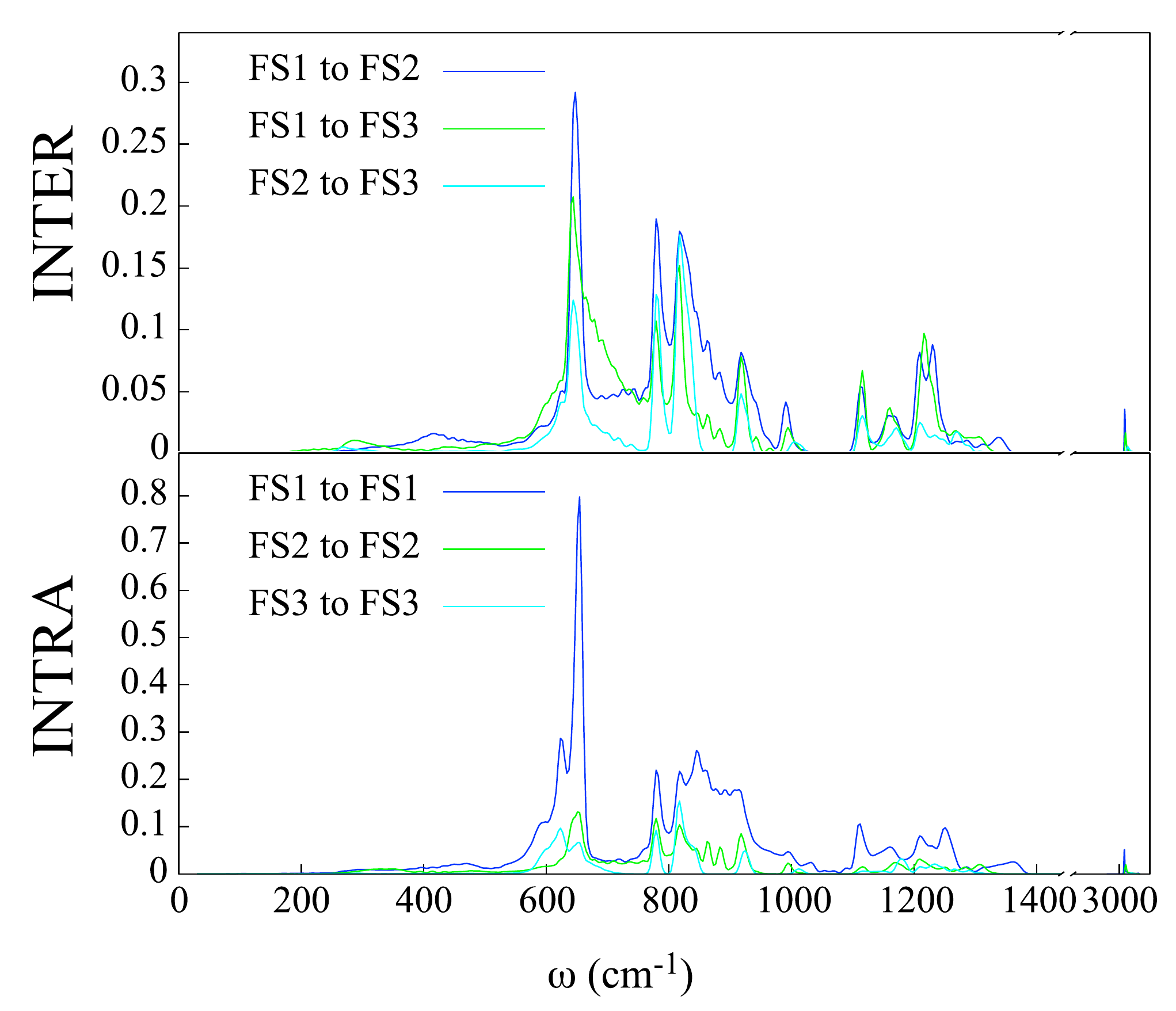}
   \caption
 {Band resolved Eliashberg spectral function for the case at the highest doping. The label FS1, FS2 and FS3 correspond the the first, second and third crossed bands respectively as defined in Fig.~\ref{fig:111_Fermi}.
   }
   \label{fig:111_a2F_band}
\end{figure}

We can now see how the various intra- and inter-band processes contributes to the final $\lambda$. This is obtained by computing the band-resolved Eliashberg spectral function (Fig.~\ref{fig:111_a2F_band}) and by decomposing the electron-phonon coupling constant over the three Fermi surfaces:

\begin{equation}
\label{eq:lambda_decomp}
[\lambda] = \mqty(\lambda_{11}&\lambda_{12}&\lambda_{13}\\\lambda_{21}&\lambda_{22}&\lambda_{23}\\\lambda_{31}&\lambda_{32}&\lambda_{33}) = \mqty(0.21&0.11&0.09\\0.11&0.06&0.05\\0.09&0.05&0.04)
\end{equation}

From this analysis it is possible to observe that the relevant processes are those concerning the first band. The intra-band scattering of the first Fermi surface (which is clearly an in-plane process) gives the highest contribution, while the other two intra-band processes are negligible. The other important processes for the superconductive phase transition are the inter-band scattering between the first and the second band (which is due to an in-plane mode) and between the first and the third bands (which is due to an out-of-plane mode). Indeed, the second and the third Fermi surfaces have similar electronic densities of states at the Fermi level and, since the electron-phonon interaction depends linearly on the DOS, they will contribute in the same way.

Finally, as the Wannier interpolation gives information on the electron-phonon matrix elements for each $\vb{q}$-point in the Brillouin zone, we can compute the following quantity:

\begin{equation}
\begin{split}
\tilde{\lambda}_{nm}(\vb{q}) =& \frac{\lambda_{nm}(\vb{q})}{N_{f}(\vb{q})}=\\
=&\frac{\sum_{\vb{k},\nu}(\abs{g_{\vb{k}n,\vb{k}+\vb{q}m}^{\nu}}^2/\omega_{q}^\nu)\delta(\epsilon_{\vb{k}n})\delta(\epsilon_{\vb{k}+\vb{q}m})}{\sum_{\vb{k}}\delta(\epsilon_{\vb{k}n})\delta(\epsilon_{\vb{k}+\vb{q}m})}
\end{split}
\end{equation}
which is the $\vb{q}$-dependent electron-phonon coupling constant normalized by the nesting factor. In this way we can investigate how the electron-phonon interaction strength varies over the Brillouin zone. Indeed, the actual $\lambda$ is an average of $\tilde{\lambda}_{nm}(\vb{q})$ over the $\vb{q}$-points of the FBZ:

\begin{equation}
\lambda=\sum_{nm}\sum_{\vb{q}}\frac{1}{N_q}\tilde{\lambda}_{nm}(\vb{q})\cdot N_{f}(\vb{q})
\end{equation}

\begin{figure*}[h]
\centering
\subfloat[]
	{ \includegraphics[width=0.8\linewidth]{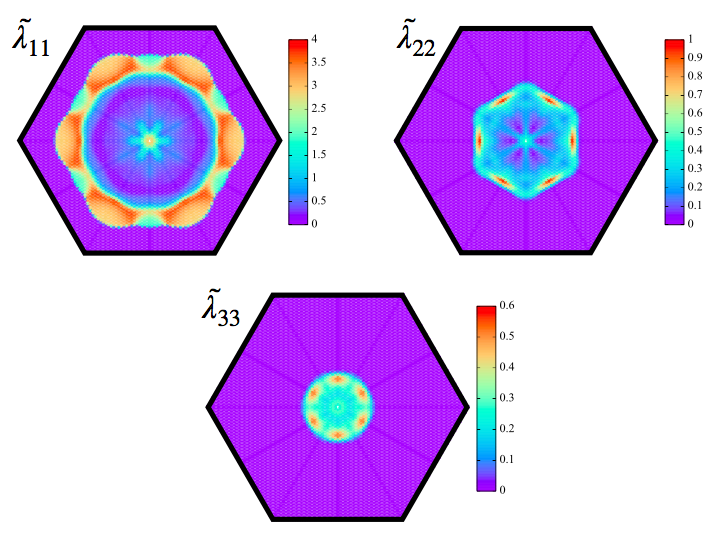}}\\
\subfloat[]
	{ \includegraphics[width=0.8\linewidth]{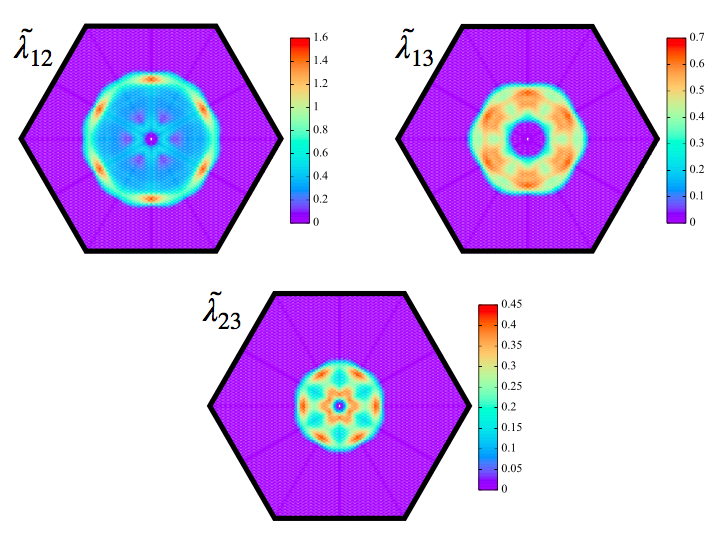}}
    \caption
 {$\vb{q}$-dependent electron-phonon coupling constant normalized by the nesting factor, resolved per band as labelled by Fig.~\ref{fig:111_Fermi}. This is a measure of how the electron-phonon interaction strength varies over the Brillouin zone.
   }
   \label{fig:111_lambda_q}
\end{figure*}

In Fig.~\ref{fig:111_lambda_q} we plot such quantity for each intra-band ($n=m$) and inter-band ($n\ne m$) process. First of all even if the regions of variation of the electron-phonon matrix elements are not circles due to the topology of the Fermi surfaces, we can discern the circle-like and annulus-like shapes which we have used in the simplified model. Nevertheless, the main assumption of constant $\abs{g_{nm,\Gamma}^\nu}^2$ inside these portions of the Brillouin zone is actually wrong: indeed high values of $\tilde{\lambda}$ can be reached, even if they are limited to very small areas of the Brillouin zone and, as a consequence, they are killed when we perform the average over the $\vb{q}$-points. This is a consequence of the topology of the electronic bands, which limits the possible $\vb{q}$ that can nest the various parts of the Fermi surfaces. Moreover it is also interesting to notice that, in most cases, it is $\abs{\vb{q}}\sim2k_{F}$ (corresponding to the wave vector responsible of the Kohn anomaly) that gives the highest values of $\tilde{\lambda}$.

\section{\label{sec:Eliashberg}Isotropic Linearized Single-Band Eliashberg Computation}

Allen-Dynes's equation (Eq.~\ref{eq:AD}) only gives a good qualitative estimate of the superconductive critical temperature but it is not quantitatively correct\cite{Carbotte}, therefore we compute $T_\text{C}$ via the Eliashberg equations\cite{Eliashberg1,Eliashberg2}. Since the electronic density of states is step-like we can assume it to be constant at the Fermi energy. Moreover the Fermi surface shown in Fig.~\ref{fig:111_Fermi}(c) can be considered to be isotropic in $\vb{k}$-space. As a consequence we numerically solve the two coupled isotropic linearized single-band Eliashberg equations, which are given on the imaginary axis\cite{AllenMitrovic} by:

\begin{equation}
\label{eq:ILE_1}
Z(i\omega_n)=1+\frac{\pi}{\beta\omega_n}\sum_{n'}\frac{\omega_{n'}}{S(i\omega_{n'})}\lambda(i\omega_n,i\omega_{n'})
\end{equation}

\begin{equation}
\label{eq:ILE_2}
Z(i\omega_n)\Delta(i\omega_n)=\frac{\pi}{\beta}\sum_{n'}\frac{\Delta(i\omega_{n'})}{S(i\omega_{n'})}\Bigl[\lambda(i\omega_n,i\omega_{n'})-\mu^*_c\Bigr]
\end{equation}
where $i\omega_n=i\frac{\pi}{\beta}(2n+1)$ are the Matsubara frequencies (for fermions), $\beta=1/k_BT$ is the inverse temperature ($k_B$ is the Boltzmann constant), $\mu^*_C$ is the Coulomb pseudopotential, $Z(i\omega_n)$ is the mass renormalization function, $\Delta(i\omega_n)$ is the superconductive energy gap and $S(i\omega_n)=\sqrt{\omega_n^2+\Delta(i\omega_n)^2}$. The superconducting critical temperature $T_C$ is defined as the temperature where the energy gap closes, i.e. when $\Delta(T_{\text{C}})=0$.

The electron-phonon coupling constant is defined as:

\begin{equation}
\lambda(i\omega_n,i\omega_{n'})=\int_0^{\omega_{max}}\frac{2\omega}{(\omega_n-\omega_m)^2+\omega^2}\alpha^2F(\omega)d\omega
\end{equation}
which in our computations was evaluated using the Eliashberg spectral function $\alpha^2F(\omega)$ obtained through the Wannier procedure described in the previous section. \\

The sums in Eq.~\ref{eq:ILE_1} and Eq.~\ref{eq:ILE_2} diverge unless one limits the number of Matsubara frequencies (or, in other words, the energy range). Usually, the energy range is upper-limited by a cutoff $\omega_c=4\omega_{\text{max}}$ or $\omega_c=10\omega_{\text{max}}$, where $\omega_{\text{max}}$ is the maximum phonon frequency after which the $\alpha^2F(\omega)$ is equal to zero. In this case, the Coulomb pseudopotential is different from zero only in an energy range $[-\omega_c,\omega_c]$, i.e. $\mu^*_c=\mu^*\cdot\theta(\omega-\abs{\omega_c})$.

Instead of defining a cutoff energy, we compute the energy gap $\Delta_0 = \Delta(i\omega_0)$ for increasing numbers of Matsubara frequencies, $n_{\mathrm{max}}$ (which corresponds to increasing the energy range) and find the value of $n_{\mathrm{max}}$ necessary  to ensure the saturation of the gap value. Then, we put to zero the Coulomb pseudopotential after a typical phonon frequency taken to be proportional to the square-root of the second moment of the Eliashberg spectral function:

\begin{equation}
\omega_{typ}=8\cdot\sqrt{\int_{0}^{\omega_{\max}}\omega^2\alpha^2F(\omega)d\omega}
\end{equation}

In our previous estimation of the superconducting critical temperature through Allen-Dynes formula, we used typical values of the Coulomb pseudopotential $\mu^*\in[0.13,0.14]$ for boron-doped diamond (as discussed in the previous section). However it is known\cite{Carbotte} that if we used the same $\mu^*$ in solving the Eliashberg equations we would get a higher value for $T_\text{C}$. Therefore we need to find a suitable value for the effective electron-electron potential which will be used for solving Eq.~\ref{eq:ILE_1} and Eq.~\ref{eq:ILE_2}. In order to do so, we solve the Eliashberg equations for boron-doped diamond at a boron concentration of $1.85\%$, where it is known experimentally that $T_\text{C}\approx4$ K. The relevant Eliashberg spectral function $\alpha^2F(\omega)$ is taken from Ref.~\onlinecite{Blase}, where it was computed via density functional perturbation theory (DFPT) using a $3\times3\times3$ supercell and one substitutional boron atom, with a corresponding electron-phonon coupling constant of $\lambda=0.43$. In order to obtain the experimental critical temperature $T_\text{C}=4$ K we have to increase the value of the Coulomb pseudopotential to $\mu^*=0.17$, which we then assume to be also representative of the hydrogenated (111)-diamond surface doped via electrochemical gating.\\

When solving the Eliashberg equations for our slab at a doping value $n_{dop,3}$, the first thing we can notice is that, for each value of the Coulomb pseudopotential, $\Delta$ converges at $T=10$ K when $n_{\mathrm{max}}=192$, i.e. when $\omega_c\approx1040$ meV. Since in our case $\omega_{max}\approx3200$ cm$^{-1}=397$ meV, the cutoff energy used in the standard approach would have been $\omega_c=4 \omega_{max}=1588$ meV.
As mentioned above, the critical temperature we get if we solve the Eliashberg equations using $\mu^*\in[0.13,0.14]$ is actually bigger than the one we obtain via Allen-Dynes's formula: as a matter of fact we get $T_{\text{C}}(\mu^*=0.13)=44.9$ K and $T_{\text{C}}(\mu^*=0.14)=42.9$ K instead of 34.93 K and 29.60 K, respectively (see Table \ref{tab:wann_vs_G}). If we use $\mu^*=0.17$, instead, the superconducting phase transition occurs at $T_C=36.3$ K. Moreover, the estimated value of the gap as $T\rightarrow0$ K is $\Delta(0)=5.89$ meV and therefore:

\begin{equation}
\frac{2\Delta(0)}{k_BT_{\text{C}}}=3.76
\end{equation}
which is close to the BCS value $3.54$.

\section{\label{sec:conlusion}Conclusions}

In this work we investigated the occurrence of field-effect induced superconductivity in the hydrogenated $(111)$ diamond surface by first-principes calculations including the effects of charging and of the intense electric field at all steps in the calculation
(electronic structure, structural and vibrational properties and electron phonon coupling). This has been possible due to the recent
methodological developments in Ref. \onlinecite{Thibault}. Previous works~\cite{Freeman,Sano} studied the hydrogenated (110) diamond surface,
however neglecting the effect of the electric field on vibrational properties and on the electron-phonon coupling. Furthermore, the electron-phonon
interaction was calculated only at zone center and assumed to be constant throughout the Brillouin zone. We have shown that this approximation
 overestimates of approximately 35$\%$ the electron-phonon interaction. Finally, it is worth to recall that the  hydrogenated (110) orientation is not the stable termination in CVD grown polycrystalline films, which instead show (100) and (111) facets. The latter orientation has been experimentally observed~\cite{Yamaguchi,Takahide} to have the higher surface capacitance ($2.6-4.6$ $\mu$F$/$cm$^2$). Thus the (111) orientation of diamond allows accumulating larger charge densities at the surface and, consequently, obtaining a higher number of carriers at the Fermi level.

Our calculations show that high $T_{\text{C}}$ superconductivity emerges at electron doping of the order of $n=6\times 10^{14}$ cm$^{-2}$. The critical temperature, calculated by using the  Allen and Dynes/McMillan equation, ranges between $29$ and $35$ K for values of $\mu^*$ typical of boron-doped diamond. The solution of the isotropic linearized single-band Eliashberg equations gives $T_{\text{C}}\simeq36$ K.
Superconductivity is mostly supported by planar vibrations and to a lesser extent by out-of-plane vibrations. The average electron-phonon coupling
is $\lambda=0.81$ and the logarithmic averaged frequency is $\omega_{\rm log}\approx 670$ cm$^{-1}$. The coupling arises partly from interband scattering, so that our values of $T_{\text{C}}$ may be an underestimation of the real $T_{\text{C}}$, as superconductivity in this system is multiband in nature. Our work demonstrates that achieving high-$T_{\text{C}}$ superconductivity in field-effect doped hydrogenated diamond is possible, even though hole charge densities of the order of $6\times10^{14}$cm$^{-2}$ are required.\\

\begin{acknowledgments}
Computational resources were provided by hpc@polito, which is a project of Academic Computing within the Department of Control and Computer Engineering at the Politecnico di Torino (http://hpc.polito.it), and by GENCI- [TGCC/CINES/IDRIS] (Grant 2019-91202). M. C. and F.M acknowledge support from the Graphene Flagship Core 2 grant number 785219. M.C. acknowledges support from Agence Nationale de la Recherche under references ANR-13-IS10-0003-01. D. R. acknowledges G. A. Ummarino for fruitful discussions.
\end{acknowledgments}

\bibliographystyle{apsrev4-1}

\end{document}